\documentclass[oneside]{article}
\usepackage{authblk}
\usepackage{xfrac}

\usepackage{csquotes}

\usepackage{mathtools}
\usepackage{graphicx}
\usepackage{esint}
\usepackage[unicode=true,
 bookmarks=false,
breaklinks=false,pdfborder={0 0 1},backref=false,colorlinks=false]
 {hyperref}
\usepackage[a4paper, total={210mm,297mm},margin=2.0cm]{geometry}

\usepackage{lineno,hyperref}
\usepackage{amsmath,bm}
\usepackage[utf8]{inputenc}
\usepackage{hyperref}
\usepackage{lmodern}
\usepackage[T1]{fontenc}
\usepackage[english]{babel}
\usepackage{graphicx}
\usepackage{float}
\usepackage{pgf,tikz}
\usepackage{pgfplots}
\usepackage{multicol}
\usepackage{color}
\usepackage{xspace}
\usepackage{caption}
\usepackage{subcaption}
\usepackage{multirow}
\usetikzlibrary{arrows}
\usetikzlibrary{spy}
\pgfplotsset{compat=newest}

\newlength{\figurewidth}
\newlength{\figureheight}

\title{A moving-grid approach for fluid-structure interaction problems with hybrid lattice Boltzmann method}

\author[1]{G. Di Ilio \thanks{Electronic address: \texttt{giovanni.diilio@unicusano.it}; Corresponding author}}
\author[1]{D. Chiappini}
\author[2]{S. Ubertini}
\author[3]{G. Bella}
\author[4]{S. Succi}

\affil[1]{University of Rome 'Niccolò Cusano', Via don Carlo Gnocchi 3, 00166, Rome, Italy}
\affil[2]{University of Tuscia, Largo dell'Università snc, 01100, Viterbo, Italy}
\affil[3]{University of Rome 'Tor Vergata', Via del Politecnico 1, 00133, Rome, Italy}
\affil[4]{Istituto Applicazioni Calcolo, CNR, Via dei Taurini 19, 00185, Rome, Italy}

\date{\today}

\begin{document}

\maketitle

\begin{abstract}
In this paper, we propose a hybrid lattice Boltzmann method (HLBM) for solving fluid-structure interaction problems. The proposed numerical approach is applied to model the flow induced by a vibrating thin lamina submerged in a viscous quiescent fluid. The hydrodynamic force exerted by the fluid on the solid body is described by means of a complex hydrodynamic function, whose real and imaginary parts are determined via parametric analysis.
Numerical results are validated by comparison with those from other numerical as well as experimental works available in literature. The proposed hybrid approach enhances the capability of lattice Boltzmann methods to solve fluid dynamic problems involving moving geometries.
\end{abstract}

\section{Introduction}
During the last decades an increasing attention has been addressed to model fluid-structure interaction (FSI) problems where the mutual actions played in between a viscous fluid and an immersed object determine the evolution of both fluid and solid motion. One of the most studied test-case is the oscillation of a lamina into a quiescent fluid, which may have practical applications in several technical branches, such as atomic force microscopy \cite{Maali,Kirstein}, sensors and actuators based on micro-mechanical oscillators \cite{Kimber2007,Kimber2009a,Castille,Hosaka}, cooling devices for electronics \cite{Kimber2009b}, underwater propulsion \cite{AureliKopman,Abdelnour,Pooley}, micro-electro-mechanical systems \cite{Batra2006,Batra2007} and energy harvesting through smart materials \cite{Cha,AureliPrince}. 
\\
Due to the large interest on this class of problems, several experiments as well as numerical studies have been carried out in order to characterize the behavior of such a system and predict the forces exerted on the structure by the fluid.
Interactions between the fluid and the structure are commonly treated in terms of added mass and hydrodynamic damping. These components may act both in-phase and out-of-phase with respect to the acceleration of the oscillating body.
\\
The works available in literature span a wide range of Reynolds numbers and oscillation amplitudes. In particular, such analyses concern very low Reynolds number (Stokes flow regime), where the inertial effects are neglected \cite{Tuck}, as well as high frequency/high oscillations regimes, where non linearities and second order effects play a significant role in force exchange process \cite{Facci}.
\\
The effects due to the oscillatory motion of objects submerged in a fluid are typically described by means of the Keulegan-Carpenter (KC) number, which is a non-dimensional quantity defined as the ratio between the oscillation amplitude and the characteristic length of the body (usually the cross-section), multiplied by 2$\pi$.
Therefore, the low Reynolds number regime corresponds to the limit of KC$\rightarrow 0$. This case has been firstly studied by Stokes \cite{Stokes} and Tuck \cite{Tuck}, who analyzed it numerically through linearized Navier-Stokes equations for small amplitude oscillations. By adopting a boundary integral formulation for the stream function, Tuck determined the numerical solution for the hydrodynamic load on rigid bodies of different geometries submerged in an unbounded fluid. Approaches similar to \cite{Tuck} were also employed in \cite{Green} to study the effects of the presence of a solid surface in the proximity of the oscillating lamina, and in \cite{DiIlio2014}, where the case of a vibrating lamina under a free-surface is studied. Under the same assumption of infinitely small amplitude oscillations, Sader \cite{Sader} presented a detailed theoretical analysis of the frequency response of a cantilever beam immersed in a viscous unbounded fluid and excited by an arbitrary driving force. In his work, he provided a mathematical practical formulation of that FSI problem, by introducing a complex hydrodynamic function, whose real and imaginary parts correspond to the added mass and the viscous damping, respectively, acting on the oscillating body.
\\
The findings obtained in all of the aforementioned studies are valid for values of KC close to zero. However, at higher KC, vortex shedding and convective effects become predominant and start to dominate the fluid flow phenomena. Therefore, for these cases, the simplified approaches based on linearized Navier-Stokes equations is no more valid. 
In order to cope with this issues, many experimental tests have been carried out and several numerical models have been proposed. Aureli and Porfiri \cite{Aureli} and Bidkar et al. \cite{Bidkar} studied the nonlinear vibrations of cantilever beams of rectangular cross-section undergoing large amplitude oscillations.
In a recent work, Aureli et al. \cite{AureliBasaran} extended the validity of the classical hydrodynamic function developed for small amplitude oscillation by providing a correction which takes into account nonlinear effects. Their results were obtained via computational analyses and validated by comparison with an ad hoc designed experimental study.
More recently, Tafuni and Sahin have presented results obtained through Smoothed Particle Hydrodynamics for the case of a lamina undergoing large amplitude oscillations in unbounded domain \cite{Tafuni2015} and under a free-surface \cite{Tafuni2013}. In particular, in \cite{Tafuni2015} the authors propose a novel formulation for the hydrodynamic function.
Some works concerning vibration of laminae in viscous fluids were also performed with lattice Boltzmann methods (LBMs) \cite{Ansumali,Succi,Succi2,Qian,Benzi,Higuera,DiIlioETH}.
\\
Focusing on LBMs, previous works have been developed within the framework of standard schemes, based on the Bhatnagar-Gross-Krook approximation (LBGK) on regular cartesian grids.
In particular, Falcucci et al. \cite{Falcucci} were the first studying this kind of problem with a LBM and they proposed an innovative refill procedure for the lattice sites in the proximity of the oscillating lamina. Although the presented method was proven to represent satisfactorily the fluid flow physics, the numerical prediction of the hydrodynamic forces slightly overestimates literature results.
Further, a combined lattice Boltzmann and finite element method for FSI problems has been proposed by De Rosis et al. \cite{DeRosis2013}. In \cite{DeRosis2014}, the same authors presented a coupled lattice Boltzmann-finite element approach with immersed boundary method (IB-LBM). They later applied such a method to the case of a thin lamina in a quiescent viscous fluid \cite{DeRosis2015}. Their results show a significant improvement, in terms of accuracy, in the prediction of the hydrodynamic load acting on the lamina, with respect to the previous lattice Boltzmann simulations.
In Shi and Sader \cite{Shi}, an ad-hoc LBM implementation was developed to analyze harmonic oscillatory Stokes flows in the frequency domain. The proposed method was validated by simulating the one-dimensional oscillatory Couette flow between two plates, the two-dimensional flow generated by an oscillating circular cylinder and the three-dimensional flow induced by an oscillating sphere.
Moreover, Colosqui et al. \cite{Colosqui} investigated on the high-frequency oscillations of electromechanical resonators operating in gaseous  media, and they obtained a good agreement with experimental observations.
\\
In order to enhance the capability of the lattice Boltzmann method to simulate FSI problems, we propose an extension of the hybrid lattice Boltzmann method \cite{DiIlio,DiIlio2018} to moving grids (MG-HLBM).
In this method, the region around the body is treated by means of a finite-volume LB scheme, while the outer domain is solved via a traditional LBGK approach on structured cartesian grid. 
\\
The benefits deriving from HLBM have been already shown in a recent work \cite{DiIlio}. Here, we explore the capabilities of such a method to deal with moving unstructured grids into a fixed Cartesian domain.
The results are compared with those available in literature, showing a good agreement and a significant improvement with respect to previous works using original LBM implementation.
Such an improvement is mainly due to two reasons. First, the proposed moving-grid approach allows an efficient mesh refinement. The local refinement, which is realized by means of an unstructured grid and is applied in the proximity of the solid body, follows the motion of the object. This leads to a computationally efficient and accurate method, since the outer fixed domain can be represented by a course mesh, while the solid geometry is described by a fine mesh. This feature of the MG-HLBM represents an advantage with respect to the numerical methods based on a fixed mesh, which are, in general, computationally expensive, since they require a dynamical re-meshing to capture accurately the flow physics around the moving object.
Second, the presence of an unstructured halo in the surrounding of the oscillating lamina allows an accurate evaluation of the forces, since computational nodes are located on the real body surface. This prevents from the implementation of error-prone interpolation strategies.
\\
The work is organized as follow. In Sec. \ref{Problem statement} the characteristic parameters for the problem under study are provided. In Sec. \ref{HLBM} we briefly recall the key points of the HLBM and the fundamental equations. In Sec. \ref{Refill procedure} the refill procedure for the lattice nodes in the proximity of the oscillating lamina is presented. In Sec. \ref{Numerical setup} we describe the numerical setup used for simulations. In Sec. \ref{Results} the results of this work are presented and discussed. Finally, in Sec. \ref{implementation}, we conduct a qualitative analysis of the numerical performance of the MG-HLBM and we discuss some basic implementation aspects, while in Sec. \ref{Conclusion} we summarize the key points of the work in the conclusions.

\section{Problem statement}
\label{Problem statement}
We consider the unsteady flow generated by a thin, rigid lamina undergoing transverse harmonic oscillations in a viscous, incompressible, quiescent fluid. The lamina is assumed to be infinitely long, such that the analysis is conducted within a two-dimensional framework.
The cross-section of the lamina is a rectangle with length \textit{L} and thickness \textit{T}, with an aspect ratio $T/L$ equal to 1/100.
At each time \textit{t} the vertical position of the lamina is given by a sinusoidal function $y(t) = Asin(\omega t)$, where \textit{A} denotes the oscillation amplitude and $\omega$ is the radian frequency.
\\
The specific problem of interest is governed by two parameters, that are the non-dimensional amplitude of oscillation $\varepsilon=A/L$ and the frequency parameter $\beta=\rho\omega L^2/(2\pi\mu)$, respectively \cite{Falcucci,Aureli,Wang}. In particular, the non-dimensional amplitude of oscillation is directly related to the KC number as $ \kappa=2\pi\varepsilon$. 
\\
The Reynolds number based on the length of the lamina \textit{L} and the maximum value of the lamina velocity, $V_{max}=A\omega$, can be expressed as a function of $\varepsilon$ and $\beta$, as follows: Re = $2\pi\beta\varepsilon$.
\\
In this work, we perform numerical simulations in the range of oscillatory Reynolds numbers Re = $2.5-94.2$, by considering the following parameters values: $\varepsilon$ = [0.02, 0.03, 0.04, 0.05, 0.075, 0.1] and $\beta$ = [20, 50, 100, 150, 200, 250, 300]. In this regime, effects of fluid inertia are not negligible, and vorticity generated by the moving boundary leads to nonlinear inertial and damping effects \cite{Bidkar,Aureli}.
In particular, the set of control parameters ($\beta$,$\varepsilon$) chosen for each simulation satisfies the following approximate correlation, provided by Aureli et al. \cite{AureliBasaran}:
\begin{equation}
\beta<2.6\varepsilon^{-1.6}.
\label{eq:betalimit}
\end{equation}
Relation \eqref{eq:betalimit} represents a reference boundary within which the force exerted by the fluid on the lamina is expected to be a purely harmonic time function.

\subsection{Hydrodynamic function extraction}
\label{force}

For a cantilever beam moving in a viscous fluid under harmonic base excitation, the hydrodynamic load per unit length due to the motion of the fluid around the beam can be expressed, in the frequency domain, as \cite{Sader}:
\begin{equation}
\hat{F}(\omega)=F_0 e^{i\phi}=\frac{\pi}{4}\rho\omega^2 L^2 \Theta(\beta,\varepsilon)A
\label{eq:force_phasor}
\end{equation}
where the superimposed hat denotes a phasor quantity, and $F_0$ and $\phi$ represent the force amplitude and the phase shift between the force and the displacement of an harmonic response, respectively.
In equation \eqref{eq:force_phasor}, $\Theta(\beta,\varepsilon)$ indicates a complex hydrodynamic function. By following \cite{Aureli,AureliBasaran}, such a quantity can be expressed as the sum of two contributions:
\begin{equation}
\Theta(\beta,\varepsilon)=\Gamma(\beta)+\Delta(\beta,\varepsilon).
\label{eq:hydro_tot}
\end{equation}
The term $\Gamma(\beta)$ corresponds to the complex hydrodynamic function proposed in \cite{Sader} for infinitely small amplitude oscillations, while $\Delta(\beta,\varepsilon)$ is the correction term for finite amplitude oscillations.
For the range of control parameters $\beta$ and $\varepsilon$ considered in this work, the following semianalytical formulations have been proven to well approximate the hydrodynamic function \cite{AureliBasaran}:
\begin{equation}
\Gamma(\beta)=1.02+2.45\beta^{-1/2}-i 2.49\beta^{-1/2}
\label{eq:hydro_gamma}
\end{equation}
\begin{equation}
\Delta(\beta,\varepsilon)=-i 0.879\beta^{3/4}\varepsilon^{2}
\label{eq:hydro_delta}
\end{equation}
where \textit{i} is the imaginary unit.
\\
To extract the hydrodynamic function components, for each simulation, that is, for each set of control parameters, we assume that the force response is purely harmonic. Therefore, in the same fashion of \cite{AureliBasaran,Falcucci}, a single harmonic sine model of the form $F(t)=F_0\sin(\omega t + \phi)$ is used for the least square fitting of the force time history. After discarding the first cycle of oscillation, we consider three cycles of oscillations to identify the values of $F_0$ and $\phi$, while the radian frequency $\omega$ is provided in the fitting model as an input. From these two quantities we then compute the real and the imaginary parts of $\Theta(\beta,\varepsilon)$.

\section{Hybrid lattice Boltzmann method}
\label{HLBM}
The moving grid approach proposed in this work is based on a lattice Boltzmann method applied to hybrid structured and unstructured grids recently proposed by the authors (HLBM) \cite{DiIlio}. For clarity of exposition, in this section the key points of such a method are recalled.
\\
The HLBM combines the standard single-time relaxation lattice Boltzmann scheme with an unstructured finite-volume lattice Boltzmann formulation \cite{Ubertini,Ubertini2,Ubertini3,ZarghamiBisc,ZarghamiDiF}.
\\
The standard scheme is applied on a uniformly spaced lattice domain while the finite-volume approach is solved on a unstructured mesh of triangular elements. The two mesh components overlap to each other, entirely covering the computational domain.
\\
As far as the standard lattice Boltzmann method is concerned, this is based on the following equation:
\begin{equation}
f_i\left(\textbf{x}+\textbf{c}_i\Delta t_\textbf{s},t+\Delta t_\textbf{s}\right)-f_i\left(\textbf{x},t\right) = -\dfrac{\Delta t_\textbf{s}}{\tau_\textbf{s}}\left[f_i\left(\textbf{x},t\right) - f_i^{eq}\left(\textbf{x},t\right)\right],
\label{eq:LBE}
\end{equation}
where $\tau_\textbf{s}$ and $\Delta t_\textbf{s}$ are the relaxation time and the time step, respectively, related to the 'structured' scheme. In equation \ref{eq:LBE}, $f_i(\textbf{x},t)$ is the distribution function, representing the probability of finding a fluid particle at position $\textbf{x}$ and time \textit{t} that is moving along the \textit{i}-th lattice direction with a discrete speed $\textbf{c}_i$. The equilibrium distribution functions $f_i^{eq}(\textbf{x},t)$ are given by the following expression:
\begin{equation}
f_i^{eq}\left(\textbf{x},t\right) = w_i\rho\left(\textbf{x},t\right) \bigg\{1+\dfrac{\textbf{c}_i\cdot\textbf{u}\left(\textbf{x},t\right)}{c_s^2} +\dfrac{\left[\textbf{c}_i\cdot\textbf{u}\left(\textbf{x},t\right)\right]^2}{2c_s^4}- \dfrac{\left[\textbf{u}\left(\textbf{x},t\right)\right]^2}{2c_s^2}\bigg\}
\label{eq:local_eq}
\end{equation}
where $c_s$ is the lattice speed of sound, the parameters $w_i$ are a set of weights normalized to unity, and $\rho\left(\textbf{x},t\right)$ and $\textbf{u}\left(\textbf{x},t\right)$ are the fluid density and velocity, respectively, which are given by the first two moments of the distribution function. It can be proven that, for an ideal incompressible fluid flow, equation \ref{eq:LBE} reproduces the Navier-Stokes equations, when pressure is $p = \rho c_s^2$ and kinematic viscosity is $\nu = c_s^2 (\tau_\textbf{s}-\frac{\Delta t_\textbf{s}}{2})$.
\\
The finite-volume lattice Boltzmann method is a cell-vertex type scheme which is expressed by the following equation, for each node \textit{P} of the mesh:
\begin{equation}
\begin{split}
f_i(P,t+\Delta t_\textbf{u}) = & f_i(P,t)+\Delta t_\textbf{u}\sum_{k=0}^{K'}S_{i k}f_i(P_k,t)+ \\
& -\frac{\Delta t_\textbf{u}}{\tau_\textbf{u}}\sum_{k=0}^{K'}C_{i k}[f_i(P_k,t)-f_i^{eq}(P_k,t)],
\end{split}
\label{eq:ulbe}
\end{equation}
where $\tau_\textbf{u}$ and $\Delta t_\textbf{u}$ are the relaxation time and the time step, respectively, related to the 'unstructured' scheme \cite{Ubertini,Ubertini2}. In equation \ref{eq:ulbe} $k=0$ denotes the pivotal node $P$ and the summations run over the nodes $P_k$ connected to $P$. The quantities $S_{i k}$ and $C_{i k}$ represent the streaming and collisional matrices of the \textit{i}-th population related to the \textit{k}-th node, respectively. The equilibrium distribution functions are defined by equation \eqref{eq:local_eq}.
For the unstructured lattice Boltzmann method the theoretical kinematic viscosity is $\nu = c_s^2 \tau_\textbf{u}$ \cite{Ubertini2}.
\\
The value of the lattice speed of sound, $c_s$, is the same in both the unstructured and the standard approaches. In this work, the $D2Q9$ model is employed, therefore $c_s=1/$$\sqrt{3}$.
\\
At each time interval, the streaming-collision process described by equation \ref{eq:LBE} is performed on the structured nodes, while equation \ref{eq:ulbe} is solved on the unstructured mesh \textit{n} times, being $\Delta t_\textbf{s} = n\Delta t_\textbf{u}$. The exchange of information between the two mesh components, in terms of distribution functions, takes place at defined interpolation nodes, by applying the following set of equations:
\begin{equation}
\tilde{f_i^{\textbf{s}}} = f_i^{eq,\textbf{u}} + \frac{2\left(\tau_\textbf{s}-\Delta t_\textbf{s}\right)}{2\tau_\textbf{s}-\Delta t_\textbf{s}} f_i^{neq,\textbf{u}}
\label{eq:postcollision2_s}
\end{equation}
\begin{equation}
\tilde{f_i^{\textbf{u}}} = f_i^{eq,\textbf{s}} + \left(1-\frac{\Delta t_\textbf{s}}{2\tau_\textbf{s}}\right) f_i^{neq,\textbf{s}}+\Delta t_\textbf{u} \sum_{k=0}^{K'} S_{i k} \left[f_{i k}^{eq,\star}+f_{i k}^{neq,\star}\right]-\frac{\Delta t_\textbf{u}}{\tau_\textbf{u}} \sum_{k=0}^{K'} C_{i k} f_{i k}^{neq,\star}
\label{eq:postcollision2_u}
\end{equation}
where superscripts \textbf{s} and \textbf{u} refer to the 'structured' and the 'unstructured' nodes, respectively, and $f_{i}^{neq}$ is the non-equilibrium distribution function.
The quantities $f_{i k}^{eq,\star}$ and $f_{i k}^{neq,\star}$ in the summation terms of equation \eqref{eq:postcollision2_u} are defined as follows:
\begin{equation}
f_{i k}^{eq,\star} = f_{i k}^{eq,\textbf{s}},
\qquad
f_{i k}^{neq,\star} = \left(1-\frac{\Delta t_\textbf{s}}{2\tau_\textbf{s}}\right)f_{i k}^{neq,\textbf{s}}
\end{equation}
or
\begin{equation}
f_{i k}^{eq,\star} = f_{i k}^{eq,\textbf{u}},
\qquad
f_{i k}^{neq,\star} = f_{i k}^{neq,\textbf{u}}
\end{equation}
depending on whether the \textit{k}-th node is an interpolation node or not, respectively. The distribution functions $f_i^{eq,\textbf{u}}$ and $f_i^{neq,\textbf{u}}$ of the right-hand side of equation \eqref{eq:postcollision2_s} and $f_i^{eq,\textbf{s}}$ and $f_i^{neq,\textbf{s}}$ of the right-hand side of equation \eqref{eq:postcollision2_u} are evaluated by interpolation procedure.
\\
Equations \eqref{eq:postcollision2_s} and \eqref{eq:postcollision2_u} represent the post-collision step for the interpolation nodes of structured and unstructured mesh, respectively. Such relations are obtained by imposing consistency of viscosity in the flow field and continuity of velocity, density and stresses across the interface.

\section{Refill procedure}
\label{Refill procedure}
\begin{figure}[H]
	\centering
	\input{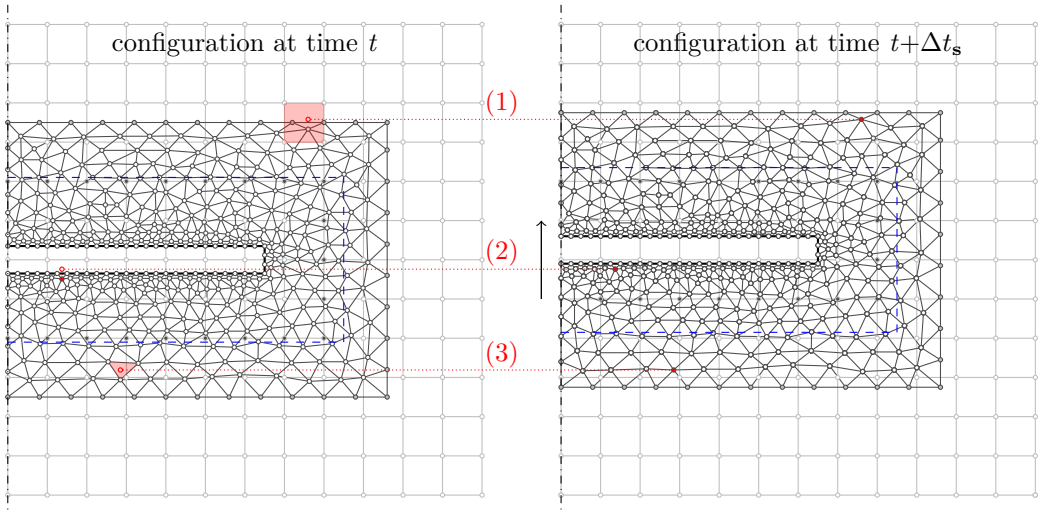}
	\caption{Schematic representation of a symmetric portion of the mesh configuration at two different time levels: \textit{t} (left) and $t + \Delta t_\textbf{s}$ (right). Dashed line is the fictitious line which is drawn in order to define the structured \textit{interpolation} nodes. Gray-filled nodes and dark gray-filled nodes represent the \textit{interpolation} points for the unstructured and the structured grid, respectively. The functional classification for the unstructured nodes is represented: \textit{type 1}, (1); \textit{type 2}, (2); \textit{type 3}, (3).
	\label{fig:refill}}
\end{figure}

In the numerical approach proposed in this work, the unstructured mesh surrounding the lamina is enabled to move within the structured fixed domain. At each time interval the hybrid method is applied and the flow field is computed all over the overlapping grid system.
To accomplish this aim, we propose a refill procedure for the unstructured lattice sites, which is performed at the beginning of each time interval. Such a procedure is based on the attribution of both macroscopic quantities and distribution functions to the relocated unstructured nodes.
\\
Figure \ref{fig:refill} illustrates a generic overlapping grid system at different times. In order to describe the procedure, we first recall the functional definition for the nodal points of each grid. According to the nomenclature adopted in \cite{DiIlio}, nodes are classified as \textit{discretization}, \textit{interpolation} and \textit{unused}.
As far as the unstructured mesh is concerned, all the boundary nodes are considered as \textit{interpolation} nodes when this boundary represents an interface with the structured mesh (gray-filled nodes in Figure \ref{fig:refill}). All of the interior unstructured nodes are considered as \textit{discretization} nodes. Regarding the structured mesh, a fictitious curve, which is defined within the unstructured grid boundaries, separates the \textit{discretization} nodes from the \textit{unused} ones. Among the structured \textit{unused} nodes, those with at least one \textit{discretization} node as neighbor, in any of the allowed lattice directions, are redefined as \textit{interpolation} nodes (dark gray-filled nodes in Figure \ref{fig:refill}).
\\
Let's then consider the unstructured mesh at a given time \textit{t}. After one time step $\Delta t_\textbf{s}$, the unstructured nodes are relocated in a new position. Since the motion of the lamina is purely translational, all nodes move with same velocity along the vertical direction.
Therefore, unstructured nodes at time $t + \Delta t_\textbf{s}$ are classified as \textit{type 1}, \textit{type 2} or \textit{type 3} depending on their location with respect to the mesh configuration at time \textit{t}. In particular, \textit{type 1} refers to those nodes which are out of the unstructured grid at level time \textit{t} and whose corresponding position is located within the physical computational domain, namely in the interior of a structured element defined by four computed nodes. \textit{Type 2} is attributed to each node lying out of the unstructured grid at level time \textit{t} and whose corresponding position is located inside the body, where no computation is performed. The nodes in the new configuration whose corresponding position at level time \textit{t} lies inside the unstructured grid is considered as \textit{type 3}. We remark that unstructured nodes defining the solid-wall of the body are not considered in such a classification. 
To ensure consistency of the fluid flow, each type of unstructured node is initialized, in the new position $\textbf{x} + \Delta \textbf{x}$, at the beginning of each structured time step, according to a specific rule.
\\
For nodes of \textit{type 1}, values of non-equilibrium distribution functions $f_i^{neq,\textbf{u}}$ are computed by means of the following equation:
\begin{equation}
f_i^{neq,\textbf{u}} = \left(1-\frac{\Delta t_\textbf{s}}{2\tau_\textbf{s}}\right) f_i^{neq,\textbf{s}}.
\label{eq:continuity_stresses}
\end{equation}
where the non-equilibrium distribution functions $f_i^{neq,\textbf{s}}$ are evaluated via a bilinear Lagrange interpolation performed over the four donor nodes constituting the structured element which encloses the unstructured node under consideration. The structured donor nodes are indeed \textit{interpolation} or \textit{discretized} nodes, thus their contributions are known. Moreover, the macroscopic variables are computed by applying the same interpolation scheme, thus allowing the reconstruction of the equilibrium distribution functions.
\\
As far as unstructured nodes of \textit{type 2} are concerned, at the beginning of each structured time step we set the same macroscopic variables and distribution functions determined for the same nodes in the previous mesh configuration. This approximation is valid as the maximum value for the velocity of the lamina is much smaller than a lattice unit (in this case is of the order of 0.01 in lattice units).
\\
For nodes of \textit{type 3}, velocity components, density and distribution functions $f_i^{\textbf{u}}$ are evaluated via barycentric interpolation by considering the three vertexes of the triangular element in the previous mesh configuration enclosing the same node at new position. Once macroscopic variables are computed, it is then possible to determine also the equilibrium distribution functions.
\\
Regarding unstructured nodes defining the solid-wall of the body, we set the velocity as a boundary condition and we impose the same density and the same distribution functions $f_i^{\textbf{u}}$ of the same node in previous mesh configuration.
\\
In addition, a dynamic redefinition of the structured \textit{interpolation} nodes is performed at each time iteration. In fact, the fictitious line designed to identify such nodes moves jointly with the unstructured grid. Therefore, for each structured \textit{interpolation} node overcome by the moving fictitious line, a new one must be defined.
\\
In order to test the refill procedure, several unstructured meshes involving all of the three described types of node have been considered. However, we emphasize that the refinement level of the unstructured meshes employed for simulating the fluid flow problem under consideration is such that the minimum distance between two nodes is greater than the distance covered by any node within one time interval. Therefore, in practice, nodes of \textit{type 2} are never defined in this particular case.

\section{Numerical setup}
\label{Numerical setup}
\begin{figure}[H]
	\centering
	\setlength\figureheight{0.60\columnwidth}
	\setlength\figurewidth{0.80\columnwidth}
	\input{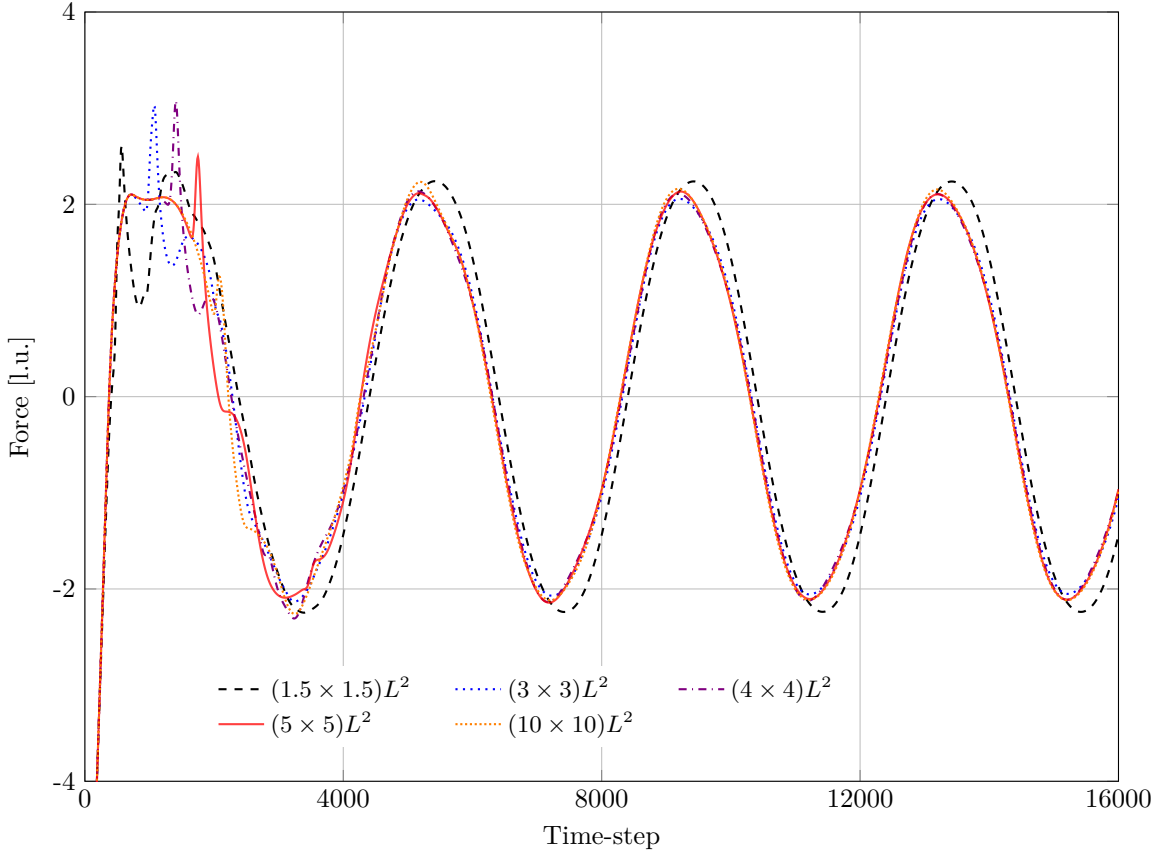}
	\caption{Hydrodynamic force time history for selected domain sizes: a) (1.5$\times$1.5)$\textit{L}^2$, black dashed line; b) (3$\times$3)$\textit{L}^2$, blue dotted line; c) (4$\times$4)$\textit{L}^2$, violet dash dotted line; d) (5$\times$5)$\textit{L}^2$, red solid line; e) (10$\times$10)$\textit{L}^2$, orange densely dotted line.}
	\label{fig:sensitivity}
\end{figure}

The computational domain is constituted by a square region inside which the lamina is defined.
No-slip wall boundary conditions are imposed at the lateral sides of the domain, while on top and bottom boundaries we set outflow conditions.
\\
In order to ensure an adequate representation of the flow field, we conduct a sensitivity analysis to select the computational domain size. Specifically, for a fixed pair of the parameters ($\beta$,$\varepsilon$), the following square domain sizes are tested: (1.5$\times$1.5)$\textit{L}^2$, (3$\times$3)$\textit{L}^2$, (4$\times$4)$\textit{L}^2$, (5$\times$5)$\textit{L}^2$ and (10$\times$10)$\textit{L}^2$. The length of the lamina is kept to a constant value. Figure \ref{fig:sensitivity} shows the time history of the resultant (non-dimensional) force exerted by the fluid on the lamina.
Although the numerical discrepancies observed by inspecting Figure \ref{fig:sensitivity} are small, the results, in terms of hydrodynamic function components vary significantly among the cases. In Table \ref{tab:sensitivity} we report the percentage error related to the case of domain size equal to (5$\times$5)$\textit{L}^2$.
\begin{table}[h!]
\centering
\begin{tabular}{ c | c | c }
Domain size  & Err \% $Re[\Theta(\beta,\varepsilon)]$ & Err \% $-Im[\Theta(\beta,\varepsilon)]$ \\ \hline
(1.5$\times$1.5)$\textit{L}^2$ & 5.53 & 53.24 \\
(3$\times$3)$\textit{L}^2$ & 3.71 & 6.94 \\
(4$\times$4)$\textit{L}^2$ & 1.58 & 1.53 \\
(5$\times$5)$\textit{L}^2$ & 0.00 & 0.00 \\
(10$\times$10)$\textit{L}^2$ & 0.35 & 0.39 \\
\end{tabular}
\caption{Effect of the domain size: relative error for real and imaginary part of the hydrodynamic function for simulation at $\beta=100$ and $\varepsilon=0.1$. The relative error represents the deviation, in percent, from the values determined with the domain size equal to (5$\times$5)$\textit{L}^2$.
\label{tab:sensitivity}}
\end{table}
The convergence study shows that the differences between the cases (5$\times$5)$\textit{L}^2$ and (10$\times$10)$\textit{L}^2$ can be considered negligible. Therefore, we select a final domain with size (5$\times$5)$\textit{L}^2$.
\\
In order to further limit the computational cost, the structured domain is subdivided in four refinement levels. The length of the lamina is fixed at 800 lattice units for all the simulations, while the relaxation time at the finest structured level is varied between 0.7 and 0.9. For this set of parameters, the maximum value of the lamina speed results to be about 0.025 lattice units, which is much below the compressibility limit.

\section{Results and discussion}
\label{Results}
In Figure \ref{fig:results} we report the numerical findings for the real and the imaginary components of the hydrodynamic function, in comparison with the semyanalitical expression (equation \ref{eq:hydro_tot}), with results obtained by Falcucci et al. \cite{Falcucci} via standard LB approach, and with those from De Rosis and L{é}v{ê}que \cite{DeRosis2015} , which use a combined IB-LBM. The results are here reported as a function of the frequency parameter $\beta$.

\begin{figure}[H]
	\centering
	\begin{subfigure}[]{0.48\textwidth}
		\centering
		\setlength\figureheight{0.60\columnwidth}
		\setlength\figurewidth{0.80\columnwidth}
		\input{./eps020.txt}
		\subcaption{}
	\end{subfigure}\hfill
	\begin{subfigure}[]{0.48\textwidth}
		\centering
		\setlength\figureheight{0.60\columnwidth}
		\setlength\figurewidth{0.80\columnwidth}
		\input{./eps030.txt}
		\subcaption{}
	\end{subfigure}
\end{figure}
\begin{figure}[H]
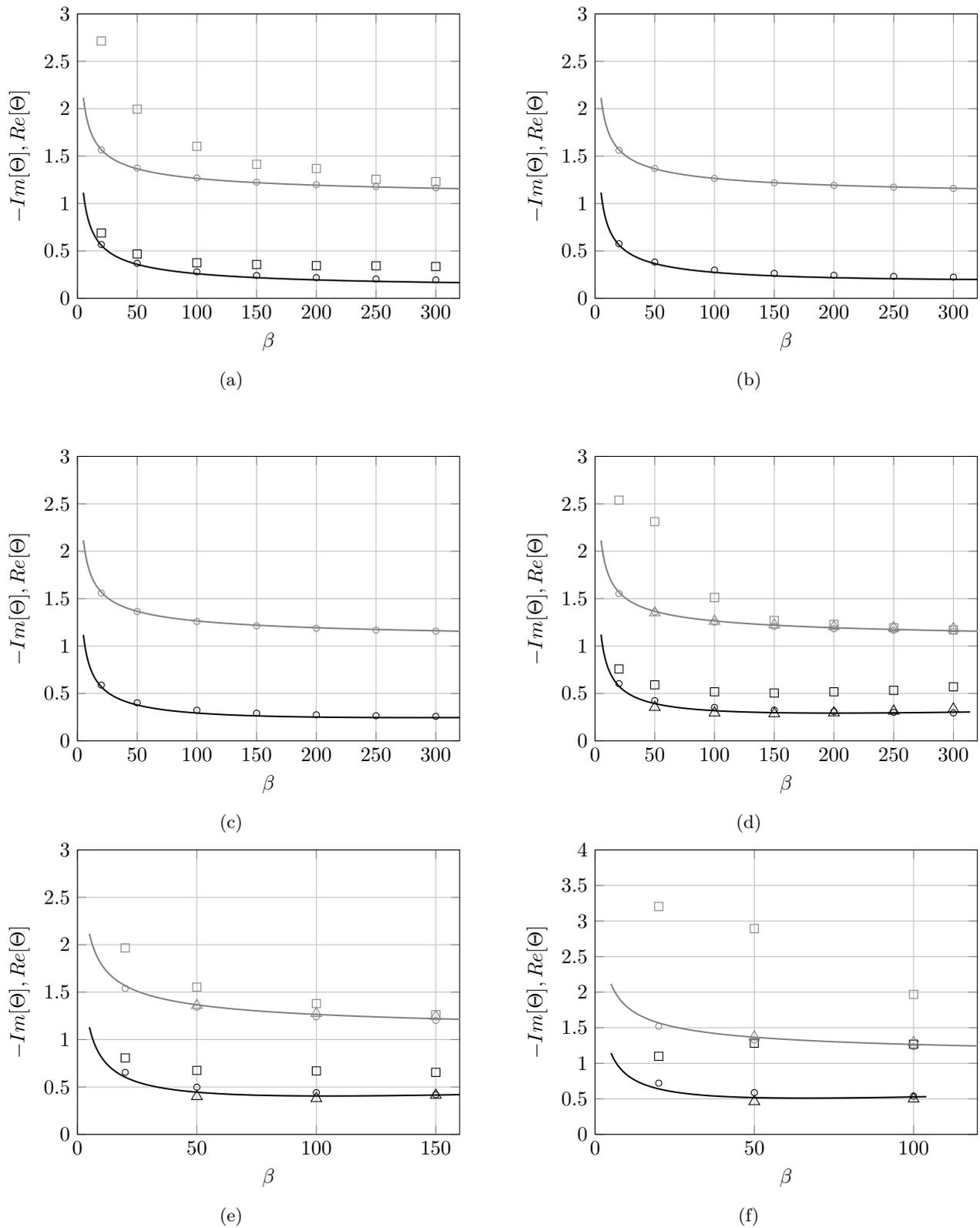
 \ContinuedFloat
	\begin{subfigure}[]{0.48\textwidth}
		\centering
		\setlength\figureheight{0.60\columnwidth}
		\setlength\figurewidth{0.80\columnwidth}
		\input{./eps040.txt}
		\subcaption{}
	\end{subfigure}\hfill
	\begin{subfigure}[]{0.48\textwidth}
		\centering
		\setlength\figureheight{0.60\columnwidth}
		\setlength\figurewidth{0.80\columnwidth}
		\input{./eps050.txt}
		\subcaption{}
	\end{subfigure}\hfill
	\begin{subfigure}[]{0.48\textwidth}
		\centering
		\setlength\figureheight{0.60\columnwidth}
		\setlength\figurewidth{0.80\columnwidth}
		\begin{tikzpicture}

\begin{axis}[%
width=\figurewidth,
height=\figureheight,
scale only axis,
xmin = 0,
xmax = 160,
xtick={0, 50, 100, 150, 200, 250, 300},
xlabel={$\beta$},
xmajorgrids,
ymin = 0,
ymax = 3,
ytick={0, 0.5, 1, 1.5, 2, 2.5, 3},
ylabel={$-Im[\Theta], Re[\Theta]$},
ymajorgrids,
legend style={draw=none, at={(0.06,0.45)},anchor=north west,legend cell align=left,font = \small}
]

\addplot [color=black, line width = 0.75pt]
  table[row sep=crcr]{%
5	1.1301	\\
6	1.0355	\\
7	0.9624	\\
8	0.9039	\\
9	0.8557	\\
10	0.8152	\\
11	0.7806	\\
12	0.7507	\\
13	0.7245	\\
14	0.7013	\\
15	0.6806	\\
16	0.6621	\\
17	0.6453	\\
18	0.6301	\\
19	0.6162	\\
20	0.6035	\\
21	0.5919	\\
22	0.5811	\\
23	0.5711	\\
24	0.5619	\\
25	0.5533	\\
26	0.5453	\\
27	0.5378	\\
28	0.5307	\\
29	0.5242	\\
30	0.5180	\\
31	0.5122	\\
32	0.5067	\\
33	0.5015	\\
34	0.4966	\\
35	0.4920	\\
36	0.4877	\\
37	0.4835	\\
38	0.4796	\\
39	0.4759	\\
40	0.4723	\\
41	0.4690	\\
42	0.4658	\\
43	0.4627	\\
44	0.4599	\\
45	0.4571	\\
46	0.4545	\\
47	0.4520	\\
48	0.4496	\\
49	0.4473	\\
50	0.4451	\\
51	0.4430	\\
52	0.4410	\\
53	0.4391	\\
54	0.4373	\\
55	0.4356	\\
56	0.4340	\\
57	0.4324	\\
58	0.4309	\\
59	0.4294	\\
60	0.4280	\\
61	0.4267	\\
62	0.4255	\\
63	0.4243	\\
64	0.4231	\\
65	0.4220	\\
66	0.4210	\\
67	0.4200	\\
68	0.4190	\\
69	0.4181	\\
70	0.4173	\\
71	0.4164	\\
72	0.4157	\\
73	0.4149	\\
74	0.4142	\\
75	0.4135	\\
76	0.4129	\\
77	0.4123	\\
78	0.4117	\\
79	0.4112	\\
80	0.4107	\\
81	0.4102	\\
82	0.4097	\\
83	0.4093	\\
84	0.4089	\\
85	0.4085	\\
86	0.4081	\\
87	0.4078	\\
88	0.4075	\\
89	0.4072	\\
90	0.4069	\\
91	0.4067	\\
92	0.4065	\\
93	0.4063	\\
94	0.4061	\\
95	0.4059	\\
96	0.4058	\\
97	0.4056	\\
98	0.4055	\\
99	0.4054	\\
100	0.4054	\\
101	0.4053	\\
102	0.4052	\\
103	0.4052	\\
104	0.4052	\\
105	0.4052	\\
106	0.4052	\\
107	0.4052	\\
108	0.4052	\\
109	0.4053	\\
110	0.4054	\\
111	0.4054	\\
112	0.4055	\\
113	0.4056	\\
114	0.4057	\\
115	0.4058	\\
116	0.4060	\\
117	0.4061	\\
118	0.4062	\\
119	0.4064	\\
120	0.4066	\\
121	0.4067	\\
122	0.4069	\\
123	0.4071	\\
124	0.4073	\\
125	0.4076	\\
126	0.4078	\\
127	0.4080	\\
128	0.4082	\\
129	0.4085	\\
130	0.4087	\\
131	0.4090	\\
132	0.4093	\\
133	0.4096	\\
134	0.4098	\\
135	0.4101	\\
136	0.4104	\\
137	0.4107	\\
138	0.4110	\\
139	0.4114	\\
140	0.4117	\\
141	0.4120	\\
142	0.4123	\\
143	0.4127	\\
144	0.4130	\\
145	0.4134	\\
146	0.4137	\\
147	0.4141	\\
148	0.4145	\\
149	0.4149	\\
150	0.4152	\\
151	0.4156	\\
152	0.4160	\\
153	0.4164	\\
154	0.4168	\\
155	0.4172	\\
156	0.4176	\\
157	0.4180	\\
158	0.4184	\\
159	0.4189	\\
160	0.4193	\\
161	0.4197	\\
162	0.4201	\\
163	0.4206	\\
164	0.4210	\\
};     

\addplot [color=black, only marks, mark size=2pt, mark=square]
  table[row sep=crcr]{%
20	0.8063	\\
50	0.6749	\\
100	0.6696	\\
150	0.6541	\\
};

\addplot [color=black, only marks, mark size=3pt, mark=triangle]
  table[row sep=crcr]{%
50	0.400	\\
100 0.379	\\
150 0.413	\\
};

\addplot [color=black, only marks, mark size=1.5pt, mark=o]
  table[row sep=crcr]{%
20	0.654030846	\\
50	0.49799542	\\
100	0.440544895	\\
150	0.41791941	\\
};

\addplot [color=gray, line width = 0.75pt]
table[row sep=crcr]{%
5	2.1157	\\
6	2.0202	\\
7	1.9460	\\
8	1.8862	\\
9	1.8367	\\
10	1.7948	\\
11	1.7587	\\
12	1.7273	\\
13	1.6995	\\
14	1.6748	\\
15	1.6526	\\
16	1.6325	\\
17	1.6142	\\
18	1.5975	\\
19	1.5821	\\
20	1.5678	\\
21	1.5546	\\
22	1.5423	\\
23	1.5309	\\
24	1.5201	\\
25	1.5100	\\
26	1.5005	\\
27	1.4915	\\
28	1.4830	\\
29	1.4750	\\
30	1.4673	\\
31	1.4600	\\
32	1.4531	\\
33	1.4465	\\
34	1.4402	\\
35	1.4341	\\
36	1.4283	\\
37	1.4228	\\
38	1.4174	\\
39	1.4123	\\
40	1.4074	\\
41	1.4026	\\
42	1.3980	\\
43	1.3936	\\
44	1.3894	\\
45	1.3852	\\
46	1.3812	\\
47	1.3774	\\
48	1.3736	\\
49	1.3700	\\
50	1.3665	\\
51	1.3631	\\
52	1.3598	\\
53	1.3565	\\
54	1.3534	\\
55	1.3504	\\
56	1.3474	\\
57	1.3445	\\
58	1.3417	\\
59	1.3390	\\
60	1.3363	\\
61	1.3337	\\
62	1.3312	\\
63	1.3287	\\
64	1.3263	\\
65	1.3239	\\
66	1.3216	\\
67	1.3193	\\
68	1.3171	\\
69	1.3149	\\
70	1.3128	\\
71	1.3108	\\
72	1.3087	\\
73	1.3068	\\
74	1.3048	\\
75	1.3029	\\
76	1.3010	\\
77	1.2992	\\
78	1.2974	\\
79	1.2956	\\
80	1.2939	\\
81	1.2922	\\
82	1.2906	\\
83	1.2889	\\
84	1.2873	\\
85	1.2857	\\
86	1.2842	\\
87	1.2827	\\
88	1.2812	\\
89	1.2797	\\
90	1.2783	\\
91	1.2768	\\
92	1.2754	\\
93	1.2741	\\
94	1.2727	\\
95	1.2714	\\
96	1.2701	\\
97	1.2688	\\
98	1.2675	\\
99	1.2662	\\
100	1.2650	\\
101	1.2638	\\
102	1.2626	\\
103	1.2614	\\
104	1.2602	\\
105	1.2591	\\
106	1.2580	\\
107	1.2569	\\
108	1.2558	\\
109	1.2547	\\
110	1.2536	\\
111	1.2525	\\
112	1.2515	\\
113	1.2505	\\
114	1.2495	\\
115	1.2485	\\
116	1.2475	\\
117	1.2465	\\
118	1.2455	\\
119	1.2446	\\
120	1.2437	\\
121	1.2427	\\
122	1.2418	\\
123	1.2409	\\
124	1.2400	\\
125	1.2391	\\
126	1.2383	\\
127	1.2374	\\
128	1.2366	\\
129	1.2357	\\
130	1.2349	\\
131	1.2341	\\
132	1.2332	\\
133	1.2324	\\
134	1.2316	\\
135	1.2309	\\
136	1.2301	\\
137	1.2293	\\
138	1.2286	\\
139	1.2278	\\
140	1.2271	\\
141	1.2263	\\
142	1.2256	\\
143	1.2249	\\
144	1.2242	\\
145	1.2235	\\
146	1.2228	\\
147	1.2221	\\
148	1.2214	\\
149	1.2207	\\
150	1.2200	\\
151	1.2194	\\
152	1.2187	\\
153	1.2181	\\
154	1.2174	\\
155	1.2168	\\
156	1.2162	\\
157	1.2155	\\
158	1.2149	\\
159	1.2143	\\
160	1.2137	\\
161	1.2131	\\
162	1.2125	\\
163	1.2119	\\
164	1.2113	\\
165	1.2107	\\
166	1.2102	\\
167	1.2096	\\
168	1.2090	\\
169	1.2085	\\
170	1.2079	\\
171	1.2074	\\
172	1.2068	\\
173	1.2063	\\
174	1.2057	\\
175	1.2052	\\
176	1.2047	\\
177	1.2042	\\
178	1.2036	\\
179	1.2031	\\
180	1.2026	\\
181	1.2021	\\
182	1.2016	\\
183	1.2011	\\
184	1.2006	\\
185	1.2001	\\
186	1.1996	\\
187	1.1992	\\
188	1.1987	\\
189	1.1982	\\
190	1.1977	\\
191	1.1973	\\
192	1.1968	\\
193	1.1964	\\
194	1.1959	\\
195	1.1954	\\
196	1.1950	\\
197	1.1946	\\
198	1.1941	\\
199	1.1937	\\
200	1.1932	\\
201	1.1928	\\
202	1.1924	\\
203	1.1920	\\
204	1.1915	\\
205	1.1911	\\
206	1.1907	\\
207	1.1903	\\
208	1.1899	\\
209	1.1895	\\
210	1.1891	\\
211	1.1887	\\
212	1.1883	\\
213	1.1879	\\
214	1.1875	\\
215	1.1871	\\
216	1.1867	\\
217	1.1863	\\
218	1.1859	\\
219	1.1856	\\
220	1.1852	\\
221	1.1848	\\
222	1.1844	\\
223	1.1841	\\
224	1.1837	\\
225	1.1833	\\
226	1.1830	\\
227	1.1826	\\
228	1.1823	\\
229	1.1819	\\
230	1.1815	\\
231	1.1812	\\
232	1.1809	\\
233	1.1805	\\
234	1.1802	\\
235	1.1798	\\
236	1.1795	\\
237	1.1791	\\
238	1.1788	\\
239	1.1785	\\
240	1.1781	\\
241	1.1778	\\
242	1.1775	\\
243	1.1772	\\
244	1.1768	\\
245	1.1765	\\
246	1.1762	\\
247	1.1759	\\
248	1.1756	\\
249	1.1753	\\
250	1.1750	\\
251	1.1746	\\
252	1.1743	\\
253	1.1740	\\
254	1.1737	\\
255	1.1734	\\
256	1.1731	\\
257	1.1728	\\
258	1.1725	\\
259	1.1722	\\
260	1.1719	\\
261	1.1717	\\
262	1.1714	\\
263	1.1711	\\
264	1.1708	\\
265	1.1705	\\
266	1.1702	\\
267	1.1699	\\
268	1.1697	\\
269	1.1694	\\
270	1.1691	\\
271	1.1688	\\
272	1.1686	\\
273	1.1683	\\
274	1.1680	\\
275	1.1677	\\
276	1.1675	\\
277	1.1672	\\
278	1.1669	\\
279	1.1667	\\
280	1.1664	\\
281	1.1662	\\
282	1.1659	\\
283	1.1656	\\
284	1.1654	\\
285	1.1651	\\
286	1.1649	\\
287	1.1646	\\
288	1.1644	\\
289	1.1641	\\
290	1.1639	\\
291	1.1636	\\
292	1.1634	\\
293	1.1631	\\
294	1.1629	\\
295	1.1626	\\
296	1.1624	\\
297	1.1622	\\
298	1.1619	\\
299	1.1617	\\
300	1.1615	\\
301	1.1612	\\
302	1.1610	\\
303	1.1607	\\
304	1.1605	\\
305	1.1603	\\
306	1.1601	\\
307	1.1598	\\
308	1.1596	\\
309	1.1594	\\
310	1.1592	\\
311	1.1589	\\
312	1.1587	\\
313	1.1585	\\
314	1.1583	\\
315	1.1580	\\
316	1.1578	\\
317	1.1576	\\
318	1.1574	\\
319	1.1572	\\
320	1.1570	\\
321	1.1567	\\
322	1.1565	\\
323	1.1563	\\
324	1.1561	\\
325	1.1559	\\
326	1.1557	\\
327	1.1555	\\
328	1.1553	\\
329	1.1551	\\
330	1.1549	\\
331	1.1547	\\
332	1.1545	\\
333	1.1543	\\
334	1.1541	\\
335	1.1539	\\
336	1.1537	\\
337	1.1535	\\
338	1.1533	\\
339	1.1531	\\
340	1.1529	\\
341	1.1527	\\
342	1.1525	\\
343	1.1523	\\
344	1.1521	\\
345	1.1519	\\
346	1.1517	\\
347	1.1515	\\
348	1.1513	\\
349	1.1511	\\
350	1.1510	\\  
};

\addplot [color=gray, only marks, mark size=2pt, mark=square]
table[row sep=crcr]{%
20	1.9662	\\
50	1.5530	\\
100	1.3797	\\
150	1.2620	\\
};

\addplot [color=gray, only marks, mark size=3pt, mark=triangle]
table[row sep=crcr]{%
50	1.361	\\
100	1.276	\\
150	1.244	\\
};

\addplot [color=gray, only marks, mark size=1.5pt, mark=o]
table[row sep=crcr]{%
20	1.5388605669	\\
50	1.3409142188	\\
100	1.2401458543	\\
150	1.200990787		\\
};

\end{axis}
\end{tikzpicture}%
		\subcaption{}
	\end{subfigure}\hfill
	\begin{subfigure}[]{0.48\textwidth}
		\centering
		\setlength\figureheight{0.60\columnwidth}
		\setlength\figurewidth{0.80\columnwidth}
		\begin{tikzpicture}

\begin{axis}[%
width=\figurewidth,
height=\figureheight,
scale only axis,
xmin = 0,
xmax = 120,
xtick={0, 50, 100, 150, 200, 250, 300},
xlabel={$\beta$},
xmajorgrids,
ymin = 0,
ymax = 4,
ytick={0, 0.5, 1, 1.5, 2, 2.5, 3, 3.5, 4},
ylabel={$-Im[\Theta], Re[\Theta]$},
ymajorgrids,
legend style={draw=none, at={(0.06,0.45)},anchor=north west,legend cell align=left,font = \small}
]

\addplot [color=black, line width = 0.75pt]
  table[row sep=crcr]{%
5	1.1430	\\
6	1.0502	\\
7	0.9790	\\
8	0.9222	\\
9	0.8757	\\
10	0.8368	\\
11	0.8039	\\
12	0.7755	\\
13	0.7508	\\
14	0.7291	\\
15	0.7099	\\
16	0.6928	\\
17	0.6775	\\
18	0.6637	\\
19	0.6512	\\
20	0.6399	\\
21	0.6296	\\
22	0.6202	\\
23	0.6115	\\
24	0.6036	\\
25	0.5963	\\
26	0.5895	\\
27	0.5833	\\
28	0.5776	\\
29	0.5722	\\
30	0.5673	\\
31	0.5627	\\
32	0.5584	\\
33	0.5545	\\
34	0.5508	\\
35	0.5474	\\
36	0.5442	\\
37	0.5412	\\
38	0.5385	\\
39	0.5359	\\
40	0.5335	\\
41	0.5313	\\
42	0.5292	\\
43	0.5273	\\
44	0.5255	\\
45	0.5239	\\
46	0.5224	\\
47	0.5210	\\
48	0.5197	\\
49	0.5185	\\
50	0.5174	\\
51	0.5164	\\
52	0.5155	\\
53	0.5147	\\
54	0.5139	\\
55	0.5133	\\
56	0.5127	\\
57	0.5122	\\
58	0.5117	\\
59	0.5113	\\
60	0.5110	\\
61	0.5107	\\
62	0.5104	\\
63	0.5103	\\
64	0.5101	\\
65	0.5101	\\
66	0.5100	\\
67	0.5100	\\
68	0.5101	\\
69	0.5102	\\
70	0.5103	\\
71	0.5105	\\
72	0.5107	\\
73	0.5110	\\
74	0.5112	\\
75	0.5115	\\
76	0.5119	\\
77	0.5122	\\
78	0.5126	\\
79	0.5131	\\
80	0.5135	\\
81	0.5140	\\
82	0.5145	\\
83	0.5150	\\
84	0.5156	\\
85	0.5161	\\
86	0.5167	\\
87	0.5174	\\
88	0.5180	\\
89	0.5186	\\
90	0.5193	\\
91	0.5200	\\
92	0.5207	\\
93	0.5214	\\
94	0.5222	\\
95	0.5229	\\
96	0.5237	\\
97	0.5245	\\
98	0.5253	\\
99	0.5261	\\
100	0.5270	\\
101	0.5278	\\
102	0.5287	\\
103	0.5295	\\
104	0.5304	\\
};

\addplot [color=black, only marks, mark size=2pt, mark=square]
  table[row sep=crcr]{%
20	1.0988	\\
50	1.2820	\\
100	1.2681	\\
};

\addplot [color=black, only marks, mark size=3pt, mark=triangle]
  table[row sep=crcr]{%
50	 0.462	\\
100  0.499	\\
};

\addplot [color=black, only marks, mark size=1.5pt, mark=o]
  table[row sep=crcr]{%
20	0.721103131	\\
50	0.587247716	\\
100	0.537684963	\\
};

\addplot [color=gray, line width = 0.75pt]
table[row sep=crcr]{%
5	2.1157	\\
6	2.0202	\\
7	1.9460	\\
8	1.8862	\\
9	1.8367	\\
10	1.7948	\\
11	1.7587	\\
12	1.7273	\\
13	1.6995	\\
14	1.6748	\\
15	1.6526	\\
16	1.6325	\\
17	1.6142	\\
18	1.5975	\\
19	1.5821	\\
20	1.5678	\\
21	1.5546	\\
22	1.5423	\\
23	1.5309	\\
24	1.5201	\\
25	1.5100	\\
26	1.5005	\\
27	1.4915	\\
28	1.4830	\\
29	1.4750	\\
30	1.4673	\\
31	1.4600	\\
32	1.4531	\\
33	1.4465	\\
34	1.4402	\\
35	1.4341	\\
36	1.4283	\\
37	1.4228	\\
38	1.4174	\\
39	1.4123	\\
40	1.4074	\\
41	1.4026	\\
42	1.3980	\\
43	1.3936	\\
44	1.3894	\\
45	1.3852	\\
46	1.3812	\\
47	1.3774	\\
48	1.3736	\\
49	1.3700	\\
50	1.3665	\\
51	1.3631	\\
52	1.3598	\\
53	1.3565	\\
54	1.3534	\\
55	1.3504	\\
56	1.3474	\\
57	1.3445	\\
58	1.3417	\\
59	1.3390	\\
60	1.3363	\\
61	1.3337	\\
62	1.3312	\\
63	1.3287	\\
64	1.3263	\\
65	1.3239	\\
66	1.3216	\\
67	1.3193	\\
68	1.3171	\\
69	1.3149	\\
70	1.3128	\\
71	1.3108	\\
72	1.3087	\\
73	1.3068	\\
74	1.3048	\\
75	1.3029	\\
76	1.3010	\\
77	1.2992	\\
78	1.2974	\\
79	1.2956	\\
80	1.2939	\\
81	1.2922	\\
82	1.2906	\\
83	1.2889	\\
84	1.2873	\\
85	1.2857	\\
86	1.2842	\\
87	1.2827	\\
88	1.2812	\\
89	1.2797	\\
90	1.2783	\\
91	1.2768	\\
92	1.2754	\\
93	1.2741	\\
94	1.2727	\\
95	1.2714	\\
96	1.2701	\\
97	1.2688	\\
98	1.2675	\\
99	1.2662	\\
100	1.2650	\\
101	1.2638	\\
102	1.2626	\\
103	1.2614	\\
104	1.2602	\\
105	1.2591	\\
106	1.2580	\\
107	1.2569	\\
108	1.2558	\\
109	1.2547	\\
110	1.2536	\\
111	1.2525	\\
112	1.2515	\\
113	1.2505	\\
114	1.2495	\\
115	1.2485	\\
116	1.2475	\\
117	1.2465	\\
118	1.2455	\\
119	1.2446	\\
120	1.2437	\\
121	1.2427	\\
122	1.2418	\\
123	1.2409	\\
124	1.2400	\\
125	1.2391	\\
126	1.2383	\\
127	1.2374	\\
128	1.2366	\\
129	1.2357	\\
130	1.2349	\\
131	1.2341	\\
132	1.2332	\\
133	1.2324	\\
134	1.2316	\\
135	1.2309	\\
136	1.2301	\\
137	1.2293	\\
138	1.2286	\\
139	1.2278	\\
140	1.2271	\\
141	1.2263	\\
142	1.2256	\\
143	1.2249	\\
144	1.2242	\\
145	1.2235	\\
146	1.2228	\\
147	1.2221	\\
148	1.2214	\\
149	1.2207	\\
150	1.2200	\\
151	1.2194	\\
152	1.2187	\\
153	1.2181	\\
154	1.2174	\\
155	1.2168	\\
156	1.2162	\\
157	1.2155	\\
158	1.2149	\\
159	1.2143	\\
160	1.2137	\\
161	1.2131	\\
162	1.2125	\\
163	1.2119	\\
164	1.2113	\\
165	1.2107	\\
166	1.2102	\\
167	1.2096	\\
168	1.2090	\\
169	1.2085	\\
170	1.2079	\\
171	1.2074	\\
172	1.2068	\\
173	1.2063	\\
174	1.2057	\\
175	1.2052	\\
176	1.2047	\\
177	1.2042	\\
178	1.2036	\\
179	1.2031	\\
180	1.2026	\\
181	1.2021	\\
182	1.2016	\\
183	1.2011	\\
184	1.2006	\\
185	1.2001	\\
186	1.1996	\\
187	1.1992	\\
188	1.1987	\\
189	1.1982	\\
190	1.1977	\\
191	1.1973	\\
192	1.1968	\\
193	1.1964	\\
194	1.1959	\\
195	1.1954	\\
196	1.1950	\\
197	1.1946	\\
198	1.1941	\\
199	1.1937	\\
200	1.1932	\\
201	1.1928	\\
202	1.1924	\\
203	1.1920	\\
204	1.1915	\\
205	1.1911	\\
206	1.1907	\\
207	1.1903	\\
208	1.1899	\\
209	1.1895	\\
210	1.1891	\\
211	1.1887	\\
212	1.1883	\\
213	1.1879	\\
214	1.1875	\\
215	1.1871	\\
216	1.1867	\\
217	1.1863	\\
218	1.1859	\\
219	1.1856	\\
220	1.1852	\\
221	1.1848	\\
222	1.1844	\\
223	1.1841	\\
224	1.1837	\\
225	1.1833	\\
226	1.1830	\\
227	1.1826	\\
228	1.1823	\\
229	1.1819	\\
230	1.1815	\\
231	1.1812	\\
232	1.1809	\\
233	1.1805	\\
234	1.1802	\\
235	1.1798	\\
236	1.1795	\\
237	1.1791	\\
238	1.1788	\\
239	1.1785	\\
240	1.1781	\\
241	1.1778	\\
242	1.1775	\\
243	1.1772	\\
244	1.1768	\\
245	1.1765	\\
246	1.1762	\\
247	1.1759	\\
248	1.1756	\\
249	1.1753	\\
250	1.1750	\\
251	1.1746	\\
252	1.1743	\\
253	1.1740	\\
254	1.1737	\\
255	1.1734	\\
256	1.1731	\\
257	1.1728	\\
258	1.1725	\\
259	1.1722	\\
260	1.1719	\\
261	1.1717	\\
262	1.1714	\\
263	1.1711	\\
264	1.1708	\\
265	1.1705	\\
266	1.1702	\\
267	1.1699	\\
268	1.1697	\\
269	1.1694	\\
270	1.1691	\\
271	1.1688	\\
272	1.1686	\\
273	1.1683	\\
274	1.1680	\\
275	1.1677	\\
276	1.1675	\\
277	1.1672	\\
278	1.1669	\\
279	1.1667	\\
280	1.1664	\\
281	1.1662	\\
282	1.1659	\\
283	1.1656	\\
284	1.1654	\\
285	1.1651	\\
286	1.1649	\\
287	1.1646	\\
288	1.1644	\\
289	1.1641	\\
290	1.1639	\\
291	1.1636	\\
292	1.1634	\\
293	1.1631	\\
294	1.1629	\\
295	1.1626	\\
296	1.1624	\\
297	1.1622	\\
298	1.1619	\\
299	1.1617	\\
300	1.1615	\\
301	1.1612	\\
302	1.1610	\\
303	1.1607	\\
304	1.1605	\\
305	1.1603	\\
306	1.1601	\\
307	1.1598	\\
308	1.1596	\\
309	1.1594	\\
310	1.1592	\\
311	1.1589	\\
312	1.1587	\\
313	1.1585	\\
314	1.1583	\\
315	1.1580	\\
316	1.1578	\\
317	1.1576	\\
318	1.1574	\\
319	1.1572	\\
320	1.1570	\\
321	1.1567	\\
322	1.1565	\\
323	1.1563	\\
324	1.1561	\\
325	1.1559	\\
326	1.1557	\\
327	1.1555	\\
328	1.1553	\\
329	1.1551	\\
330	1.1549	\\
331	1.1547	\\
332	1.1545	\\
333	1.1543	\\
334	1.1541	\\
335	1.1539	\\
336	1.1537	\\
337	1.1535	\\
338	1.1533	\\
339	1.1531	\\
340	1.1529	\\
341	1.1527	\\
342	1.1525	\\
343	1.1523	\\
344	1.1521	\\
345	1.1519	\\
346	1.1517	\\
347	1.1515	\\
348	1.1513	\\
349	1.1511	\\
350	1.1510	\\  
};

\addplot [color=gray, only marks, mark size=2pt, mark=square]
table[row sep=crcr]{%
20	3.2049	\\
50	2.8930	\\
100	1.9681	\\
};

\addplot [color=gray, only marks, mark size=3pt, mark=triangle]
table[row sep=crcr]{%
50	1.372 \\
100	1.293 \\
};

\addplot [color=gray, only marks, mark size=1.5pt, mark=o]
table[row sep=crcr]{%
20	1.5209697728 \\
50	1.3253261073 \\
100	1.23291163	 \\
};

\end{axis}
\end{tikzpicture}%
		\subcaption{}
	\end{subfigure}
	\caption{Real part $Re[\Theta(\beta,\varepsilon)]$ (gray) and imaginary part $-Im[\Theta(\beta,\varepsilon)]$ (black) of the complex hydrodynamic function. The results determined via MG-HLBM (circles) are compared with the semianalytical formula \ref{eq:hydro_tot} provided by Aureli et al. \cite{AureliBasaran} (lines), results from De Rosis and L{é}v{ê}que \cite{DeRosis2015} (triangles, panels (d),(e),(f)), and the numerical findings from Falcucci et al. \cite{Falcucci} (squares, panels (a),(d),(e),(f)). (a): $\varepsilon = 0.02$; (b): $\varepsilon = 0.03$; (c): $\varepsilon = 0.04$; (d): $\varepsilon = 0.05$; (e): $\varepsilon = 0.075$; (f): $\varepsilon = 0.1$.}
	\label{fig:results}
\end{figure}

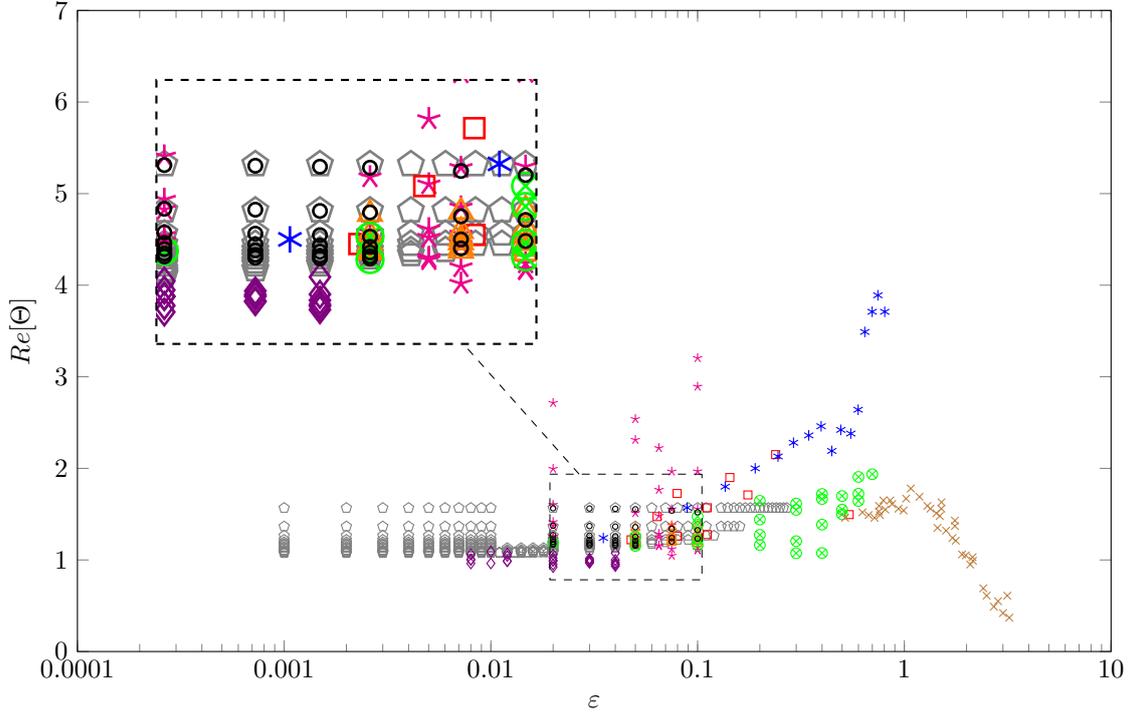
\begin{figure}[H]
	\centering 
	\setlength\figureheight{0.50\columnwidth}
	\setlength\figurewidth{0.80\columnwidth}
	\begin{tikzpicture}[spy using outlines]

\begin{axis}[%
width=\figurewidth,
height=\figureheight,
scale only axis,
xmode=log,
xmin = 0.0001,
xmax = 10,
xtick={0.0001, 0.001, 0.01, 0.1, 1, 10},
xticklabels={0.0001, 0.001, 0.01, 0.1, 1, 10},
xlabel={$\varepsilon$},
ymin = 0,
ymax = 7,
ytick={0, 1, 2, 3, 4, 5, 6, 7},
yticklabels={0, 1, 2, 3, 4, 5, 6, 7},
ylabel={$Re[\Theta]$},
legend style={draw=none, at={(0.03,0.95)},anchor=north west,legend cell align=left,font = \small}
]

\addplot [color=gray, mark size=2pt, only marks, mark=pentagon]
  table[row sep=crcr]{
0.001	1.568	\\
0.001	1.366	\\
0.001	1.265	\\
0.001	1.220	\\
0.001	1.193	\\
0.001	1.175	\\
0.001	1.161	\\
0.001	1.143	\\
0.001	1.130	\\
0.001	1.109	\\
0.001	1.097	\\
0.001	1.089	\\
0.001	1.083	\\
0.001	1.079	\\
0.001	1.075	\\
0.002	1.568	\\
0.002	1.366	\\
0.002	1.265	\\
0.002	1.220	\\
0.002	1.193	\\
0.002	1.175	\\
0.002	1.161	\\
0.002	1.143	\\
0.002	1.130	\\
0.002	1.109	\\
0.002	1.097	\\
0.002	1.089	\\
0.002	1.083	\\
0.002	1.079	\\
0.002	1.075	\\
0.003	1.568	\\
0.003	1.366	\\
0.003	1.265	\\
0.003	1.220	\\
0.003	1.193	\\
0.003	1.175	\\
0.003	1.161	\\
0.003	1.143	\\
0.003	1.130	\\
0.003	1.109	\\
0.003	1.097	\\
0.003	1.089	\\
0.003	1.083	\\
0.003	1.079	\\
0.003	1.075	\\
0.004	1.568	\\
0.004	1.366	\\
0.004	1.265	\\
0.004	1.220	\\
0.004	1.193	\\
0.004	1.175	\\
0.004	1.161	\\
0.004	1.143	\\
0.004	1.130	\\
0.004	1.109	\\
0.004	1.097	\\
0.004	1.089	\\
0.004	1.083	\\
0.004	1.079	\\
0.004	1.075	\\
0.005	1.568	\\
0.005	1.366	\\
0.005	1.265	\\
0.005	1.220	\\
0.005	1.193	\\
0.005	1.175	\\
0.005	1.161	\\
0.005	1.143	\\
0.005	1.130	\\
0.005	1.109	\\
0.005	1.097	\\
0.005	1.089	\\
0.005	1.083	\\
0.005	1.079	\\
0.005	1.075	\\
0.006	1.568	\\
0.006	1.366	\\
0.006	1.265	\\
0.006	1.220	\\
0.006	1.193	\\
0.006	1.175	\\
0.006	1.161	\\
0.006	1.143	\\
0.006	1.130	\\
0.006	1.109	\\
0.006	1.097	\\
0.006	1.089	\\
0.006	1.083	\\
0.006	1.079	\\
0.006	1.075	\\
0.007	1.568	\\
0.007	1.366	\\
0.007	1.265	\\
0.007	1.220	\\
0.007	1.193	\\
0.007	1.175	\\
0.007	1.161	\\
0.007	1.143	\\
0.007	1.130	\\
0.007	1.109	\\
0.007	1.097	\\
0.007	1.089	\\
0.007	1.083	\\
0.007	1.079	\\
0.007	1.075	\\
0.008	1.568	\\
0.008	1.366	\\
0.008	1.265	\\
0.008	1.220	\\
0.008	1.193	\\
0.008	1.175	\\
0.008	1.161	\\
0.008	1.143	\\
0.008	1.130	\\
0.008	1.109	\\
0.008	1.097	\\
0.008	1.089	\\
0.008	1.083	\\
0.008	1.079	\\
0.008	1.075	\\
0.009	1.568	\\
0.009	1.366	\\
0.009	1.265	\\
0.009	1.220	\\
0.009	1.193	\\
0.009	1.175	\\
0.009	1.161	\\
0.009	1.143	\\
0.009	1.130	\\
0.009	1.109	\\
0.009	1.097	\\
0.009	1.089	\\
0.009	1.083	\\
0.009	1.079	\\
0.009	1.075	\\
0.01	1.568	\\
0.01	1.366	\\
0.01	1.265	\\
0.01	1.220	\\
0.01	1.193	\\
0.01	1.175	\\
0.01	1.161	\\
0.01	1.143	\\
0.01	1.130	\\
0.01	1.109	\\
0.01	1.097	\\
0.01	1.089	\\
0.01	1.083	\\
0.01	1.079	\\
0.01	1.075	\\
0.011	1.083	\\
0.011	1.079	\\
0.011	1.075	\\
0.012	1.130	\\
0.012	1.109	\\
0.012	1.097	\\
0.012	1.089	\\
0.012	1.083	\\
0.012	1.079	\\
0.012	1.075	\\
0.013	1.083	\\
0.013	1.079	\\
0.013	1.075	\\
0.014	1.130	\\
0.014	1.109	\\
0.014	1.097	\\
0.014	1.089	\\
0.014	1.083	\\
0.014	1.079	\\
0.014	1.075	\\
0.015	1.083	\\
0.015	1.079	\\
0.016	1.130	\\
0.016	1.109	\\
0.016	1.097	\\
0.016	1.089	\\
0.016	1.083	\\
0.016	1.079	\\
0.017	1.083	\\
0.018	1.130	\\
0.018	1.109	\\
0.018	1.097	\\
0.018	1.089	\\
0.018	1.083	\\
0.019	1.083	\\
0.02	1.568	\\
0.02	1.366	\\
0.02	1.265	\\
0.02	1.220	\\
0.02	1.193	\\
0.02	1.175	\\
0.02	1.161	\\
0.02	1.143	\\
0.02	1.130	\\
0.02	1.109	\\
0.02	1.097	\\
0.02	1.089	\\
0.02	1.083	\\
0.03	1.568	\\
0.03	1.366	\\
0.03	1.265	\\
0.03	1.220	\\
0.03	1.193	\\
0.03	1.175	\\
0.03	1.161	\\
0.03	1.143	\\
0.03	1.130	\\
0.03	1.109	\\
0.04	1.568	\\
0.04	1.366	\\
0.04	1.265	\\
0.04	1.220	\\
0.04	1.193	\\
0.04	1.175	\\
0.04	1.161	\\
0.04	1.143	\\
0.04	1.130	\\
0.05	1.568	\\
0.05	1.366	\\
0.05	1.265	\\
0.05	1.220	\\
0.05	1.193	\\
0.05	1.175	\\
0.05	1.161	\\
0.06	1.568	\\
0.06	1.366	\\
0.06	1.265	\\
0.06	1.220	\\
0.06	1.193	\\
0.06	1.175	\\
0.07	1.568	\\
0.07	1.366	\\
0.07	1.265	\\
0.07	1.220	\\
0.07	1.193	\\
0.08	1.568	\\
0.08	1.366	\\
0.08	1.265	\\
0.08	1.220	\\
0.09	1.568	\\
0.09	1.366	\\
0.09	1.265	\\
0.09	1.220	\\
0.1		1.568	\\
0.1		1.366	\\
0.1		1.265	\\
0.11	1.568	\\
0.11	1.366	\\
0.11	1.265	\\
0.12	1.265	\\
0.13	1.568	\\
0.13	1.366	\\
0.14	1.568	\\
0.14	1.366	\\
0.15	1.568	\\
0.15	1.366	\\
0.16	1.568	\\
0.16	1.366	\\
0.17	1.568	\\
0.18	1.568	\\
0.19	1.568	\\
0.2		1.568	\\
0.21	1.568	\\
0.22	1.568	\\
0.23	1.568	\\
0.24	1.568	\\
0.25	1.568	\\
0.26	1.568	\\
0.27	1.568	\\
};

\addplot [color=magenta, mark size=1.8pt, only marks, mark=star]
  table[row sep=crcr]{
0.02	2.715 \\
0.02	1.995 \\
0.02	1.603 \\
0.02	1.414 \\
0.02	1.370 \\
0.02	1.256 \\
0.02	1.231 \\
0.05	2.539 \\
0.05	2.312 \\
0.05	1.512 \\
0.05	1.269 \\
0.05	1.228 \\
0.05	1.192 \\
0.05	1.170 \\
0.065	2.222 \\
0.065	1.766 \\
0.065	1.481 \\
0.065	1.285 \\
0.065	1.248 \\
0.065	1.160 \\
0.065	1.150 \\
0.075	1.966 \\
0.075	1.553 \\
0.075	1.380 \\
0.075	1.262 \\
0.075	1.217 \\
0.075	1.047 \\
0.075	1.119 \\
0.1		3.205 \\
0.1		2.893 \\
0.1		1.968 \\
0.1		1.558 \\
0.1		1.136 \\
0.1		1.106 \\
0.1		1.109 \\
};

\addplot [color=red, mark size=1.5pt, only marks, mark=square]
  table[row sep=crcr]{
0.0796	1.726 \\
0.1432	1.900 \\
0.2387	2.150 \\
0.0637	1.473 \\
0.1114	1.570 \\
0.1751	1.710 \\
0.0477	1.220 \\
0.0796	1.260 \\
0.1114	1.277 \\
0.5411	1.495 \\
};

\addplot [color=blue, mark size=2pt, only marks, mark=asterisk]
  table[row sep=crcr]{
0.035	1.24 \\
0.089	1.57 \\
0.136	1.80 \\
0.19	2.00 \\
0.245	2.13 \\
0.291	2.28 \\
0.345	2.36 \\
0.396	2.46 \\
0.446	2.19 \\
0.493	2.42 \\
0.551	2.38 \\
0.598	2.64 \\
0.644	3.49 \\
0.699	3.71 \\
0.745	3.89 \\
0.804	3.71 \\
};

\addplot [color=orange, mark size=2pt, only marks, mark=triangle]
  table[row sep=crcr]{
0.05	1.351	\\
0.05	1.261	\\
0.05	1.224	\\
0.05	1.205	\\
0.05	1.193	\\
0.05	1.184	\\
0.075	1.361	\\
0.075	1.276	\\
0.075	1.244	\\
0.075	1.227	\\
0.075	1.205	\\
0.075	1.190	\\
0.1	1.372	\\
0.1	1.293	\\
0.1	1.255	\\
0.1	1.224	\\
0.1	1.199	\\
0.1	1.179	\\
};

\addplot [color=violet, mark size=2pt, only marks, mark=diamond]
  table[row sep=crcr]{
0.01	1.091	\\
0.02	0.951	\\
0.03	0.972	\\
0.04	1.076	\\
0.01	0.962	\\
0.02	1.058	\\
0.03	0.968	\\
0.04	0.949	\\
0.008	1.005	\\
0.012	1.043	\\
0.02	0.991	\\
0.03	1.015	\\
0.04	0.956	\\
0.008	0.962	\\
0.012	0.987	\\
0.02	0.923	\\
0.03	1.000	\\
0.04	1.000	\\
0.008	1.000	\\
0.012	0.987	\\
0.02	1.019	\\
0.03	0.994	\\
0.04	0.974	\\
0.008	1.062	\\
0.012	1.069	\\
0.02	0.991	\\
0.03	0.991	\\
0.04	0.933	\\
};

\addplot [color=green, mark size=2pt, only marks, mark=otimes]
  table[row sep=crcr]{
0.1		1.4736	\\
0.2		1.6457	\\
0.3		1.6159	\\
0.4		1.6654	\\
0.5		1.5499	\\
0.6		1.6437	\\
0.7		1.9357	\\
0.1		1.3839	\\
0.2		1.4413	\\
0.3		1.5461	\\
0.4		1.7207	\\
0.5		1.6978	\\
0.6		1.9063	\\
0.05	1.2536	\\
0.1		1.2235	\\
0.2		1.2740	\\
0.3		1.2049	\\
0.4		1.3857	\\
0.5		1.4982	\\
0.6		1.7232	\\
0.02	1.1896	\\
0.05	1.1517	\\
0.1		1.1630	\\
0.2		1.1623	\\
0.3		1.0735	\\
0.4		1.0754	\\
};

\addplot [color=brown, mark size=2pt, only marks, mark=x]
table[row sep=crcr]{
0.517	1.46	\\
0.627	1.52	\\
0.672	1.49	\\
0.718	1.46	\\
0.763	1.49	\\
0.754	1.53	\\
0.809	1.55	\\
0.755	1.59	\\
0.792	1.65	\\
0.855	1.65	\\
0.873	1.61	\\
0.937	1.56	\\
0.992	1.54	\\
1.02	1.67	\\
1.075	1.78	\\
1.184	1.69	\\
1.293	1.59	\\
1.338	1.56	\\
1.429	1.51	\\
1.512	1.63	\\
1.511	1.56	\\
1.484	1.48	\\
1.447	1.35	\\
1.593	1.32	\\
1.757	1.37	\\
1.757	1.43	\\
1.711	1.24	\\
1.774	1.21	\\
1.902	1.06	\\
1.984	1.05	\\
2.084	0.95	\\
2.148	0.99	\\
2.139	1.03	\\
2.411	0.69	\\
2.502	0.61	\\
2.711	0.49	\\
2.839	0.55	\\
3.002	0.42	\\
3.149	0.61	\\
3.221	0.37	\\
};

\addplot [color=black, mark size=1pt, only marks, mark=o]
table[row sep=crcr]{
0.02	1.5646	\\
0.03	1.5617	\\
0.04	1.5578	\\
0.05	1.5534	\\
0.075	1.5389	\\
0.1		1.5210	\\
0.02	1.3740	\\
0.03	1.3699	\\
0.04	1.3644	\\
0.05	1.3580	\\
0.075	1.3409	\\
0.1		1.3253	\\
0.02	1.2707	\\
0.03	1.2652	\\
0.04	1.2588	\\
0.05	1.2525	\\
0.075	1.2401	\\
0.1		1.2329	\\
0.02	1.2248	\\
0.03	1.2188	\\
0.04	1.2127	\\
0.05	1.2076	\\
0.075	1.2010	\\
0.02	1.1979	\\
0.03	1.1918	\\
0.04	1.1865	\\
0.05	1.1830	\\
0.02	1.1785	\\
0.03	1.1726	\\
0.04	1.1684	\\
0.05	1.1665	\\
0.02	1.1655	\\
0.03	1.1601	\\
0.04	1.1570	\\
0.05	1.1563	\\
};

\coordinate (spypoint) at (0.045, 1.36);
\coordinate (magnifyglass) at (0.002, 4.8);
\spy[black,draw,dashed,height=3.5cm,width=5cm,magnification=2.5,connect spies] on (spypoint) in node[fill=white] at (magnifyglass);

\end{axis}
\end{tikzpicture}%
	\caption{Aggregated results for the real part $Re[\Theta(\beta,\varepsilon)]$ of the complex hydrodynamic function (color online). Black circles: present study; gray pentagons: \cite{AureliBasaran}; magenta stars: \cite{Falcucci}; red squared: \cite{Aureli}; blue asterisks: \cite{Bidkar}; orange triangles: \cite{DeRosis2015}; violet diamonds: \cite{Jalalisendi}; green crosshairs: \cite{Tafuni2015}; brown crosses: \cite{Graham,Singh}}
	\label{fig:aggregate_re}
\end{figure}

\begin{figure}[H]
	\centering 
	\setlength\figureheight{0.50\columnwidth}
	\setlength\figurewidth{0.80\columnwidth}
	\begin{tikzpicture}[spy using outlines]

\begin{axis}[%
width=\figurewidth,
height=\figureheight,
scale only axis,
xmode=log,
xmin = 0.0001,
xmax = 10,
xtick={0.0001, 0.001, 0.01, 0.1, 1, 10},
xticklabels={0.0001, 0.001, 0.01, 0.1, 1, 10},
xlabel={$\varepsilon$},
ymin = 0,
ymax = 8,
ytick={0, 1, 2, 3, 4, 5, 6, 7, 8},
yticklabels={0, 1, 2, 3, 4, 5, 6, 7, 8},
ylabel={$-Im[\Theta]$},
legend style={draw=none, at={(0.03,0.95)},anchor=north west,legend cell align=left,font = \small}
]

\addplot [color=gray, mark size=2pt, only marks, mark=pentagon]
  table[row sep=crcr]{
0.001	0.557	\\
0.001	0.352	\\
0.001	0.249	\\
0.001	0.203	\\
0.001	0.176	\\
0.001	0.158	\\
0.001	0.144	\\
0.001	0.125	\\
0.001	0.111	\\
0.001	0.091	\\
0.001	0.079	\\
0.001	0.071	\\
0.001	0.065	\\
0.001	0.060	\\
0.001	0.056	\\
0.002	0.557	\\
0.002	0.352	\\
0.002	0.249	\\
0.002	0.203	\\
0.002	0.176	\\
0.002	0.158	\\
0.002	0.144	\\
0.002	0.125	\\
0.002	0.112	\\
0.002	0.091	\\
0.002	0.079	\\
0.002	0.071	\\
0.002	0.065	\\
0.002	0.060	\\
0.002	0.057	\\
0.003	0.557	\\
0.003	0.352	\\
0.003	0.249	\\
0.003	0.204	\\
0.003	0.176	\\
0.003	0.158	\\
0.003	0.144	\\
0.003	0.125	\\
0.003	0.112	\\
0.003	0.092	\\
0.003	0.080	\\
0.003	0.072	\\
0.003	0.066	\\
0.003	0.062	\\
0.003	0.058	\\
0.004	0.557	\\
0.004	0.352	\\
0.004	0.249	\\
0.004	0.204	\\
0.004	0.177	\\
0.004	0.158	\\
0.004	0.145	\\
0.004	0.126	\\
0.004	0.113	\\
0.004	0.093	\\
0.004	0.081	\\
0.004	0.073	\\
0.004	0.068	\\
0.004	0.063	\\
0.004	0.060	\\
0.005	0.557	\\
0.005	0.353	\\
0.005	0.250	\\
0.005	0.204	\\
0.005	0.177	\\
0.005	0.159	\\
0.005	0.145	\\
0.005	0.126	\\
0.005	0.114	\\
0.005	0.094	\\
0.005	0.083	\\
0.005	0.075	\\
0.005	0.070	\\
0.005	0.065	\\
0.005	0.062	\\
0.006	0.557	\\
0.006	0.353	\\
0.006	0.250	\\
0.006	0.205	\\
0.006	0.178	\\
0.006	0.159	\\
0.006	0.146	\\
0.006	0.127	\\
0.006	0.115	\\
0.006	0.095	\\
0.006	0.084	\\
0.006	0.077	\\
0.006	0.072	\\
0.006	0.068	\\
0.006	0.065	\\
0.007	0.557	\\
0.007	0.353	\\
0.007	0.250	\\
0.007	0.205	\\
0.007	0.178	\\
0.007	0.160	\\
0.007	0.147	\\
0.007	0.128	\\
0.007	0.116	\\
0.007	0.097	\\
0.007	0.086	\\
0.007	0.079	\\
0.007	0.075	\\
0.007	0.071	\\
0.007	0.069	\\
0.008	0.557	\\
0.008	0.353	\\
0.008	0.251	\\
0.008	0.206	\\
0.008	0.179	\\
0.008	0.161	\\
0.008	0.148	\\
0.008	0.130	\\
0.008	0.117	\\
0.008	0.099	\\
0.008	0.089	\\
0.008	0.082	\\
0.008	0.078	\\
0.008	0.075	\\
0.008	0.073	\\
0.009	0.557	\\
0.009	0.353	\\
0.009	0.251	\\
0.009	0.206	\\
0.009	0.180	\\
0.009	0.162	\\
0.009	0.149	\\
0.009	0.131	\\
0.009	0.119	\\
0.009	0.101	\\
0.009	0.091	\\
0.009	0.085	\\
0.009	0.081	\\
0.009	0.079	\\
0.009	0.077	\\
0.01	0.558	\\
0.01	0.354	\\
0.01	0.252	\\
0.01	0.207	\\
0.01	0.181	\\
0.01	0.163	\\
0.01	0.150	\\
0.01	0.132	\\
0.01	0.121	\\
0.01	0.104	\\
0.01	0.094	\\
0.01	0.089	\\
0.01	0.085	\\
0.01	0.083	\\
0.01	0.082	\\
0.011	0.090	\\
0.011	0.088	\\
0.011	0.087	\\
0.012	0.125	\\
0.012	0.109	\\
0.012	0.101	\\
0.012	0.097	\\
0.012	0.095	\\
0.012	0.094	\\
0.012	0.094	\\
0.013	0.100	\\
0.013	0.100	\\
0.013	0.100	\\
0.014	0.130	\\
0.014	0.116	\\
0.014	0.109	\\
0.014	0.107	\\
0.014	0.106	\\
0.014	0.106	\\
0.014	0.107	\\
0.015	0.112	\\
0.015	0.113	\\
0.016	0.135	\\
0.016	0.123	\\
0.016	0.119	\\
0.016	0.118	\\
0.016	0.119	\\
0.016	0.120	\\
0.017	0.126	\\
0.018	0.141	\\
0.018	0.132	\\
0.018	0.129	\\
0.018	0.130	\\
0.018	0.133	\\
0.019	0.141	\\
0.02	0.560	\\
0.02	0.359	\\
0.02	0.260	\\
0.02	0.218	\\
0.02	0.195	\\
0.02	0.180	\\
0.02	0.169	\\
0.02	0.156	\\
0.02	0.149	\\
0.02	0.141	\\
0.02	0.141	\\
0.02	0.144	\\
0.02	0.149	\\
0.03	0.564	\\
0.03	0.367	\\
0.03	0.274	\\
0.03	0.237	\\
0.03	0.218	\\
0.03	0.207	\\
0.03	0.201	\\
0.03	0.195	\\
0.03	0.195	\\
0.03	0.204	\\
0.04	0.570	\\
0.04	0.379	\\
0.04	0.293	\\
0.04	0.264	\\
0.04	0.251	\\
0.04	0.246	\\
0.04	0.245	\\
0.04	0.250	\\
0.04	0.260	\\
0.05	0.578	\\
0.05	0.393	\\
0.05	0.318	\\
0.05	0.297	\\
0.05	0.293	\\
0.05	0.296	\\
0.05	0.302	\\
0.06	0.587	\\
0.06	0.412	\\
0.06	0.349	\\
0.06	0.339	\\
0.06	0.344	\\
0.06	0.356	\\
0.07	0.598	\\
0.07	0.433	\\
0.07	0.385	\\
0.07	0.388	\\
0.07	0.405	\\
0.08	0.610	\\
0.08	0.458	\\
0.08	0.427	\\
0.08	0.444	\\
0.09	0.624	\\
0.09	0.486	\\
0.09	0.474	\\
0.09	0.508	\\
0.1		0.640	\\
0.1		0.517	\\
0.1		0.527	\\
0.11	0.657	\\
0.11	0.552	\\
0.11	0.585	\\
0.12	0.649	\\
0.13	0.697	\\
0.13	0.631	\\
0.14	0.720	\\
0.14	0.676	\\
0.15	0.744	\\
0.15	0.724	\\
0.16	0.770	\\
0.16	0.775	\\
0.17	0.797	\\
0.18	0.826	\\
0.19	0.857	\\
0.2		0.889	\\
0.21	0.923	\\
0.22	0.959	\\
0.23	0.997	\\
0.24	1.036	\\
0.25	1.076	\\
0.26	1.119	\\
0.27	1.163	\\
};

\addplot [color=magenta, mark size=1.8pt, only marks, mark=star]
  table[row sep=crcr]{
0.02	0.690	\\
0.02	0.469	\\
0.02	0.375	\\
0.02	0.357	\\
0.02	0.345	\\
0.02	0.343	\\
0.02	0.336	\\
0.05	0.758	\\
0.05	0.591	\\
0.05	0.517	\\
0.05	0.504	\\
0.05	0.517	\\
0.05	0.533	\\
0.05	0.570	\\
0.065	0.797	\\
0.065	0.631	\\
0.065	0.599	\\
0.065	0.627	\\
0.065	0.649	\\
0.065	0.683	\\
0.065	0.733	\\
0.075	0.806	\\
0.075	0.675	\\
0.075	0.670	\\
0.075	0.654	\\
0.075	0.680	\\
0.075	1.118	\\
0.075	0.702	\\
0.1		1.099	\\
0.1		1.282	\\
0.1		1.268	\\
0.1		0.902	\\
0.1		0.933	\\
0.1		0.852	\\
0.1		1.353	\\
};

\addplot [color=red, mark size=1.5pt, only marks, mark=square]
  table[row sep=crcr]{
0.0796	0.463	\\
0.1432	0.585	\\
0.2387	0.950	\\
0.0637	0.441	\\
0.1114	0.538	\\
0.1751	0.680	\\
0.0477	0.270	\\
0.0796	0.362	\\
0.1114	0.427	\\
0.5411	1.306	\\
};

\addplot [color=blue, mark size=2pt, only marks, mark=asterisk]
  table[row sep=crcr]{
0.035	0.205	\\
0.089	0.743	\\
0.136	0.821	\\
0.19	1.26	\\
0.245	1.49	\\
0.291	1.65	\\
0.345	1.95	\\
0.396	2.04	\\
0.446	2.58	\\
0.493	3.04	\\
0.551	3.26	\\
0.598	3.56	\\
0.644	4.49	\\
0.699	4.62	\\
0.745	4.1		\\
0.804	4.73	\\
};

\addplot [color=orange, mark size=2pt, only marks, mark=triangle]
  table[row sep=crcr]{
0.05	0.354	\\
0.05	0.295	\\
0.05	0.287	\\
0.05	0.296	\\
0.05	0.314	\\
0.05	0.337	\\
0.075	0.400	\\
0.075	0.379	\\
0.075	0.413	\\
0.075	0.462	\\
0.075	0.510	\\
0.075	0.550	\\
0.1		0.462	\\
0.1		0.499	\\
0.1		0.583	\\
0.1		0.655	\\
0.1		0.706	\\
0.1		0.742	\\
};

\addplot [color=violet, mark size=2pt, only marks, mark=diamond]
  table[row sep=crcr]{
0.01	0.027	\\
0.02	0.163	\\
0.03	0.088	\\
0.04	0.198	\\
0.01	0.058	\\
0.02	0.128	\\
0.03	0.141	\\
0.04	0.147	\\
0.008	0.025	\\
0.012	0.037	\\
0.02	0.081	\\
0.03	0.073	\\
0.04	0.193	\\
0.008	0.147	\\
0.012	0.141	\\
0.02	0.026	\\
0.03	0.083	\\
0.04	0.154	\\
0.008	0.083	\\
0.012	0.051	\\
0.02	0.077	\\
0.03	0.109	\\
0.04	0.192	\\
0.008	0.187	\\
0.012	0.155	\\
0.02	0.135	\\
0.03	0.187	\\
0.04	0.219	\\
};

\addplot [color=green, mark size=2pt, only marks, mark=otimes]
  table[row sep=crcr]{
0.1		0.6511	\\
0.2		0.9932	\\
0.3		1.2478	\\
0.4		1.5819	\\
0.5		1.9281	\\
0.6		2.3297	\\
0.7		2.9308	\\
0.1		0.7265	\\
0.2		1.1004	\\
0.3		1.5437	\\
0.4		2.2046	\\
0.5		2.7283	\\
0.6		3.2398	\\
0.05	0.4713	\\
0.1		0.7586	\\
0.2		1.2371	\\
0.3		1.7476	\\
0.4		2.4484	\\
0.5		3.1448	\\
0.6		3.9280	\\
0.02	0.2079	\\
0.05	0.5102	\\
0.1		0.8334	\\
0.2		1.3999	\\
0.3		1.9911	\\
0.4		2.6719	\\
};

\addplot [color=brown, mark size=2pt, only marks, mark=x]
table[row sep=crcr]{
0.510	1.61	\\
0.620	1.81	\\
0.639	1.72	\\
0.675	1.77	\\
0.730	1.88	\\
0.738	1.99	\\
0.775	2.04	\\
0.838	2.15	\\
0.884	2.21	\\
0.938	2.37	\\
0.957	2.18	\\
1.011	2.39	\\
1.184	2.66	\\
1.311	2.91	\\
1.275	2.65	\\
1.285	2.59	\\
1.457	3.07	\\
1.511	2.96	\\
1.602	3.29	\\
1.412	2.69	\\
1.448	2.70	\\
1.457	2.65	\\
1.594	2.79	\\
1.748	3.20	\\
1.712	2.94	\\
1.739	2.85	\\
1.857	3.00	\\
1.957	3.13	\\
2.075	3.20	\\
2.075	3.65	\\
2.457	3.66	\\
2.701	4.22	\\
2.692	4.37	\\
2.774	4.46	\\
3.020	4.25	\\
3.410	5.35	\\
3.410	5.02	\\
3.455	5.20	\\
4.127	5.23	\\
4.145	5.51	\\
4.190	6.38	\\
};

\addplot [color=black, mark size=1pt, only marks, mark=o]
table[row sep=crcr]{
0.02	0.5681	\\
0.03	0.5765	\\
0.04	0.5882	\\
0.05	0.6033	\\
0.075	0.6540	\\
0.1		0.7211	\\
0.02	0.3694	\\
0.03	0.3829	\\
0.04	0.4014	\\
0.05	0.4244	\\
0.075	0.4980	\\
0.1		0.5872	\\
0.02	0.2798	\\
0.03	0.2983	\\
0.04	0.3231	\\
0.05	0.3528	\\
0.075	0.4405	\\
0.1		0.5377	\\
0.02	0.2411	\\
0.03	0.2633	\\
0.04	0.2919	\\
0.05	0.3251	\\
0.075	0.4179	\\
0.02	0.2182	\\
0.03	0.2431	\\
0.04	0.2745	\\
0.05	0.3097	\\
0.02	0.2044	\\
0.03	0.2316	\\
0.04	0.2648	\\
0.05	0.3010	\\
0.02	0.1947	\\
0.03	0.2238	\\
0.04	0.2583	\\
0.05	0.2951	\\
};

\coordinate (spypoint) at (0.045, 0.72);
\coordinate (magnifyglass) at (0.002, 5.5);
\spy[black,draw,dashed,height=3.5cm,width=5cm,magnification=2.5,connect spies] on (spypoint) in node[fill=white] at (magnifyglass);

\end{axis}
\end{tikzpicture}%
	\caption{Aggregated results for the imaginary part $-Im[\Theta(\beta,\varepsilon)]$ of the complex hydrodynamic function (color online). Black circles: present study; gray pentagons: \cite{AureliBasaran}; magenta stars: \cite{Falcucci}; red squared: \cite{Aureli}; blue asterisks: \cite{Bidkar}; orange triangles: \cite{DeRosis2015}; violet diamonds: \cite{Jalalisendi}; green crosshairs: \cite{Tafuni2015}; brown crosses: \cite{Graham,Singh}.}
	\label{fig:aggregate_im}
\end{figure}
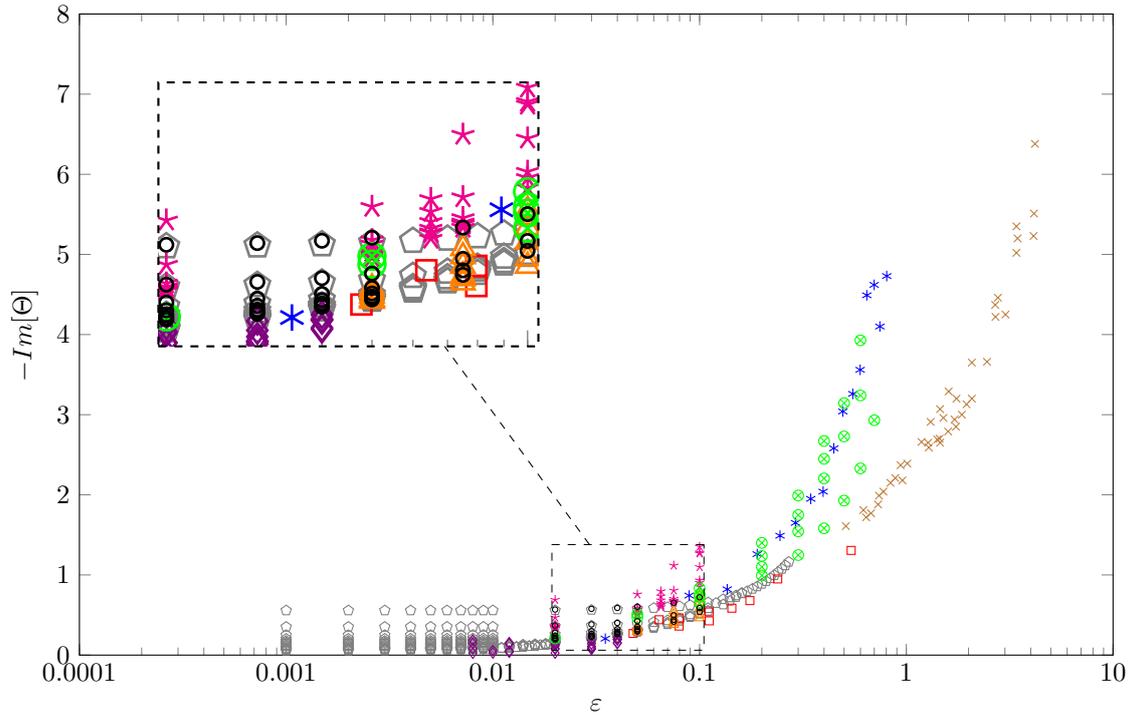

Overall, the results well predict the expected trends for both real and imaginary parts, capturing the added mass and hydrodynamic damping effects exerted by the fluid on the lamina in the range of control parameters considered. In particular, the real part $Re[\Theta(\beta,\varepsilon)]$ is nearly constant with the variations of parameter $\varepsilon$, while the imaginary part $-Im[\Theta(\beta,\varepsilon)]$ increases whit $\varepsilon$ for each fixed value of $\beta$, as suggested by the reference literature \cite{AureliBasaran}.
\\
Our results are in line with those from the semianalytical expression and with results from \cite{DeRosis2015}. We observe that the general dependence of viscous damping and added mass on the control parameters is properly estimated by the present MG-HLBM.
On the other hand, Figure \ref{fig:results} show significant differences between the MG-HLBM results and those obtained via standard LBM, which appear to be overestimated.
In spite of an increasing of the level of complexity introduced by the presence of the unstructured grid, the MG-HLBM results to be more accurate than the standard LBM in the prediction of the hydrodynamic load for the case under consideration. A crucial aspect is represented by the direct computation of the forces at the real boundary of the solid body, which is not allowed in the standard LBM. Also, the presence of an unstructured grid allows an high and flexible level of refinement in proximity of the solid body, thus leading to a reduction of the computational cost with respect to the case of uniform grid, as demonstrated in \cite{DiIlio,DiIlio2018}.
\\
For a further qualitative comparison between MG-HLBM and literature results, we report a complete overview of experimental and numerical findings in aggregated form in Figures \ref{fig:aggregate_re} and \ref{fig:aggregate_im}, for a broad range of control parameters. The real and the imaginary parts of the hydrodynamic function are reported as functions of the non-dimensional amplitude of oscillation $\varepsilon$.


\begin{figure}[H]
\centering
{\includegraphics[width = 0.5\columnwidth]{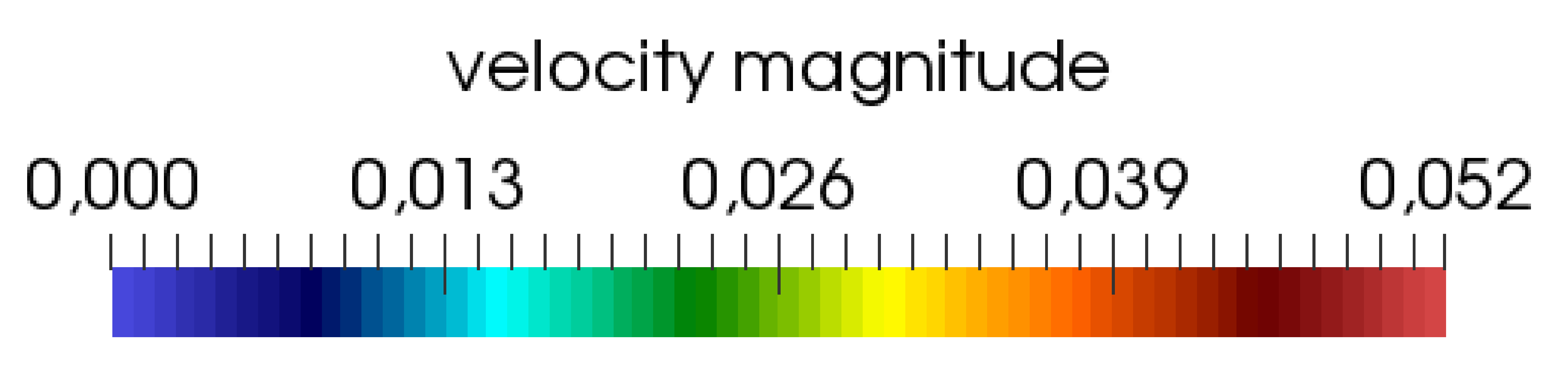}}
\\
\subcaptionbox{}
{\includegraphics[width = 0.32\columnwidth]{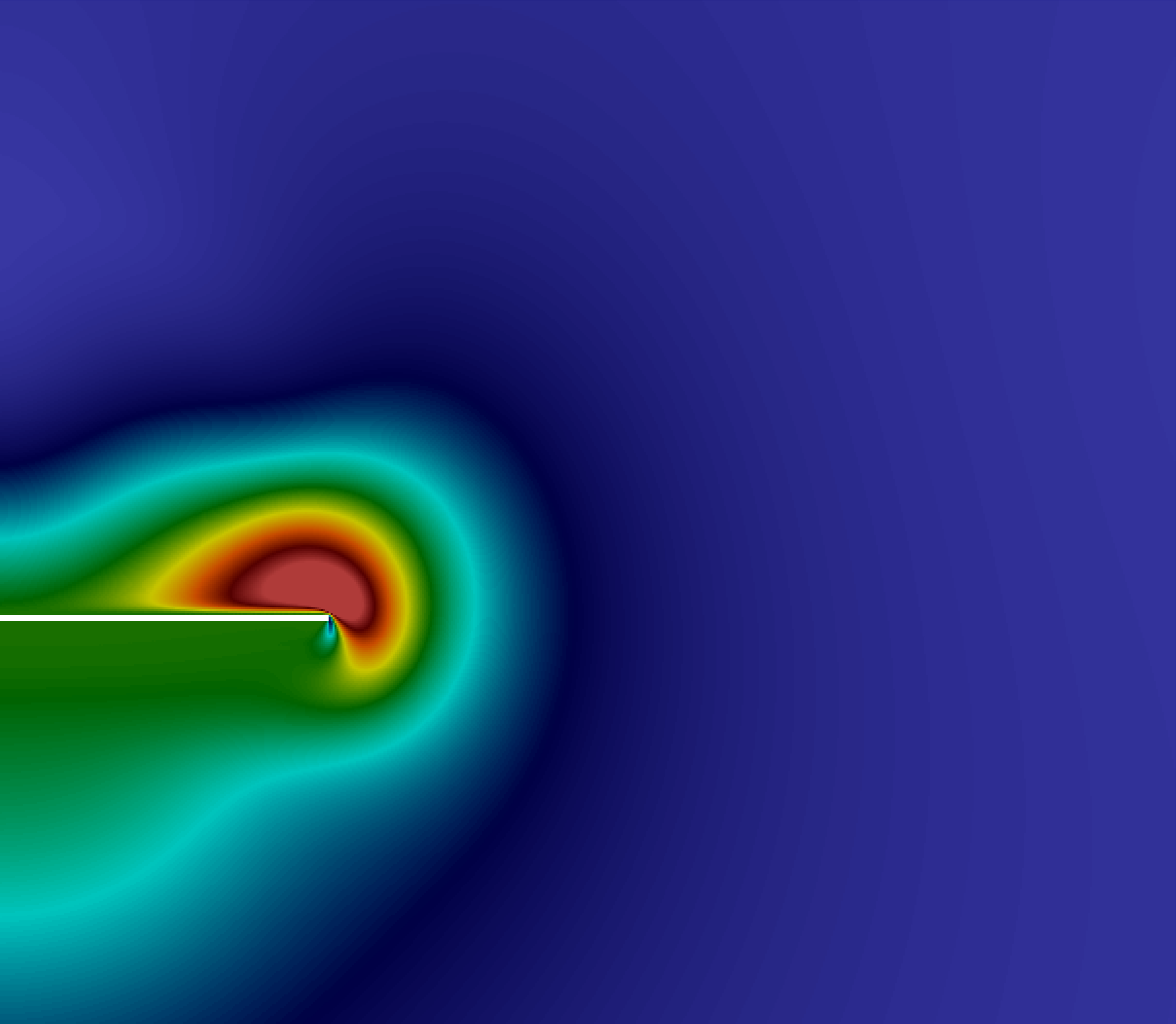}}
\subcaptionbox{}
{\includegraphics[width = 0.32\columnwidth]{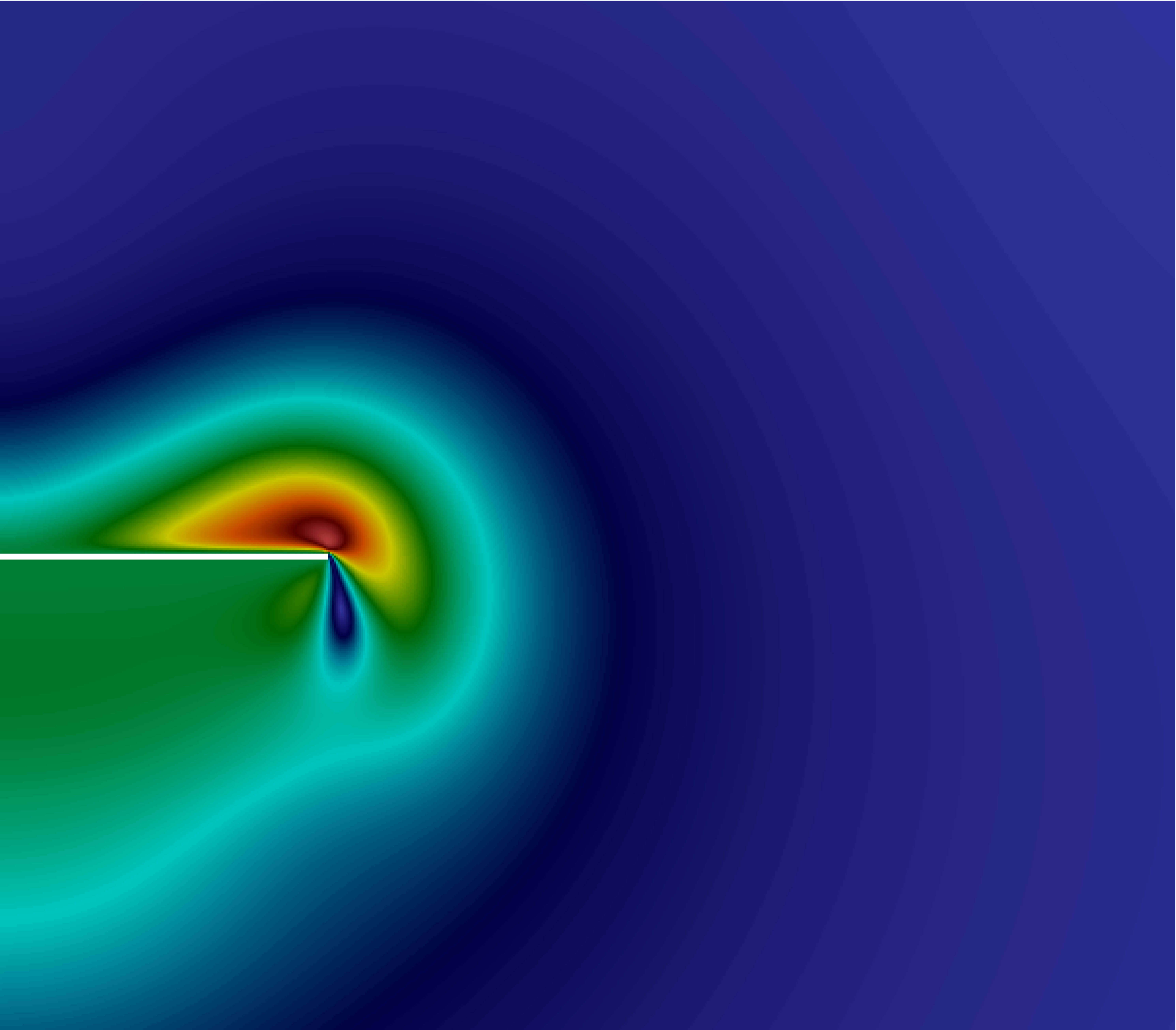}}
\subcaptionbox{}
{\includegraphics[width = 0.32\columnwidth]{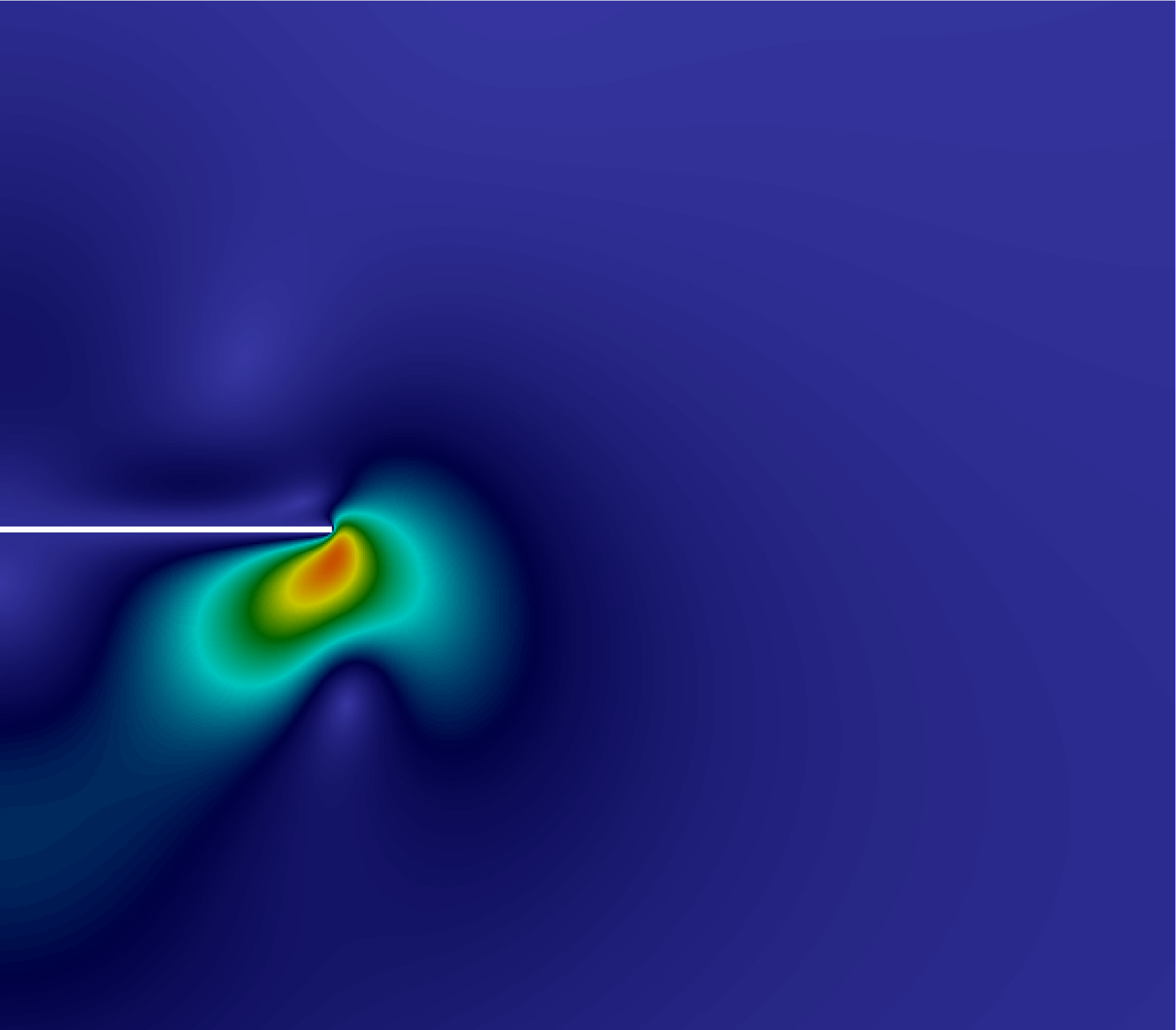}}
\\
{\includegraphics[width = 0.5\columnwidth]{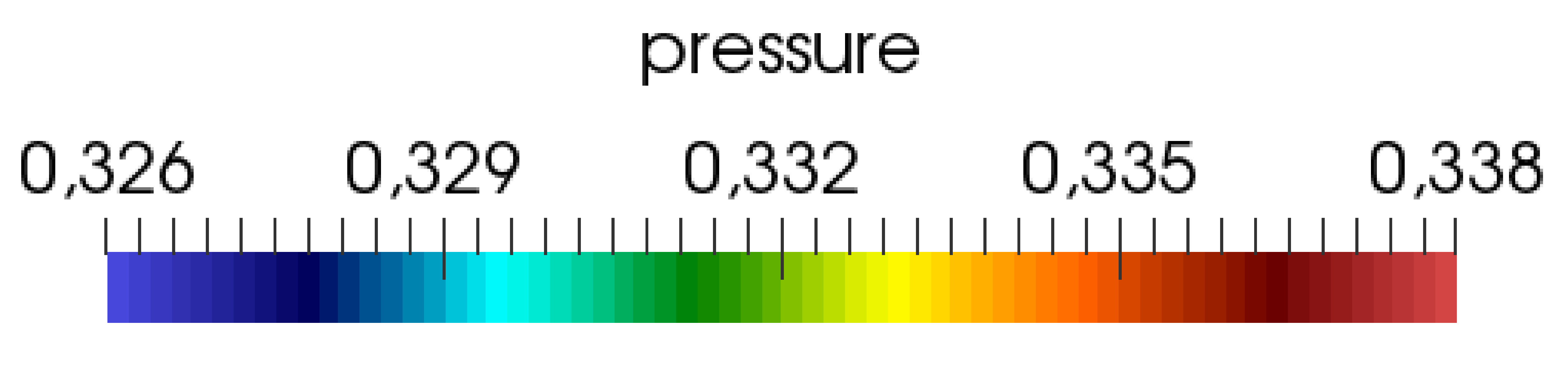}}
\\
\subcaptionbox{}
{\includegraphics[width = 0.32\columnwidth]{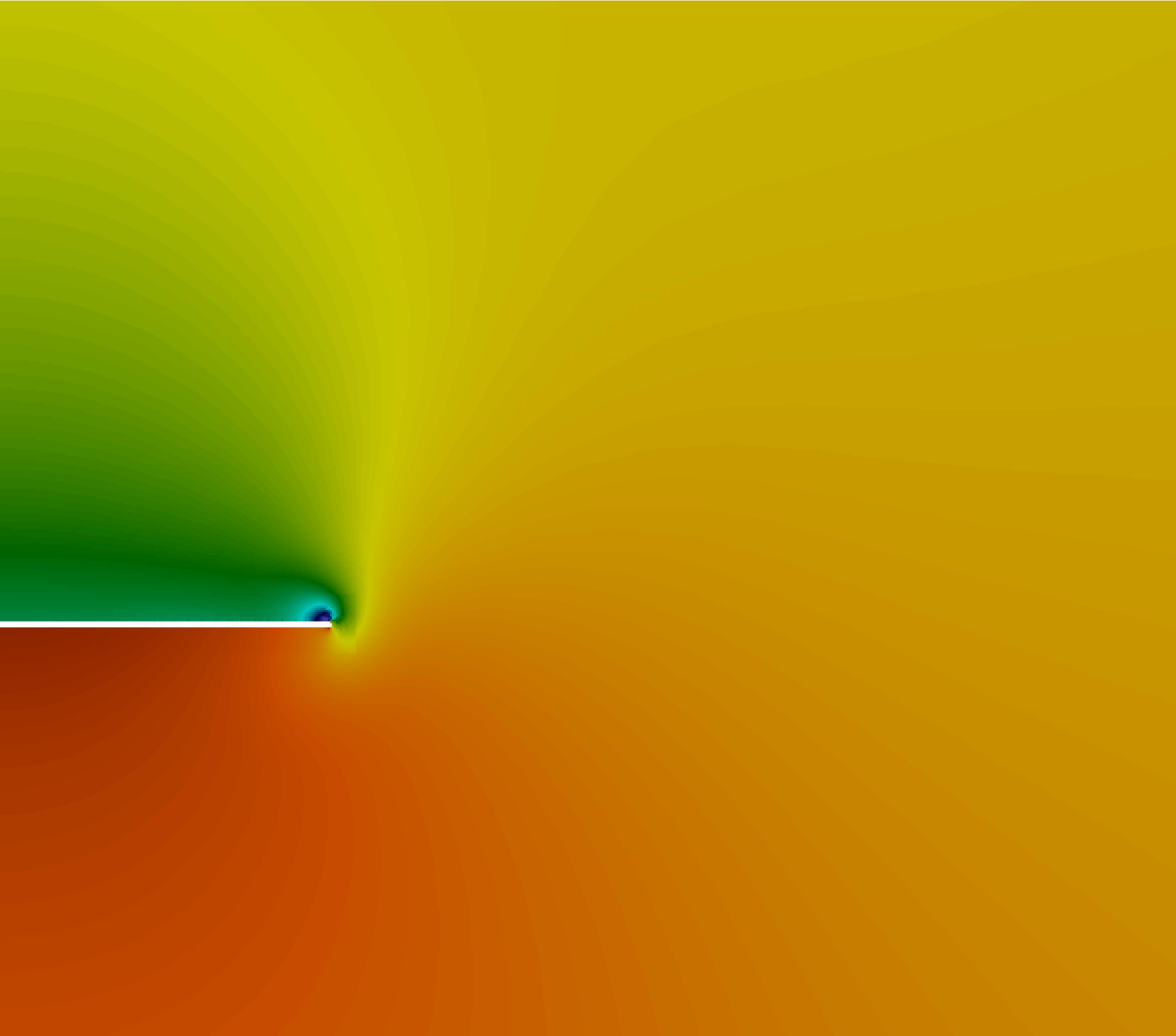}}
\subcaptionbox{}
{\includegraphics[width = 0.32\columnwidth]{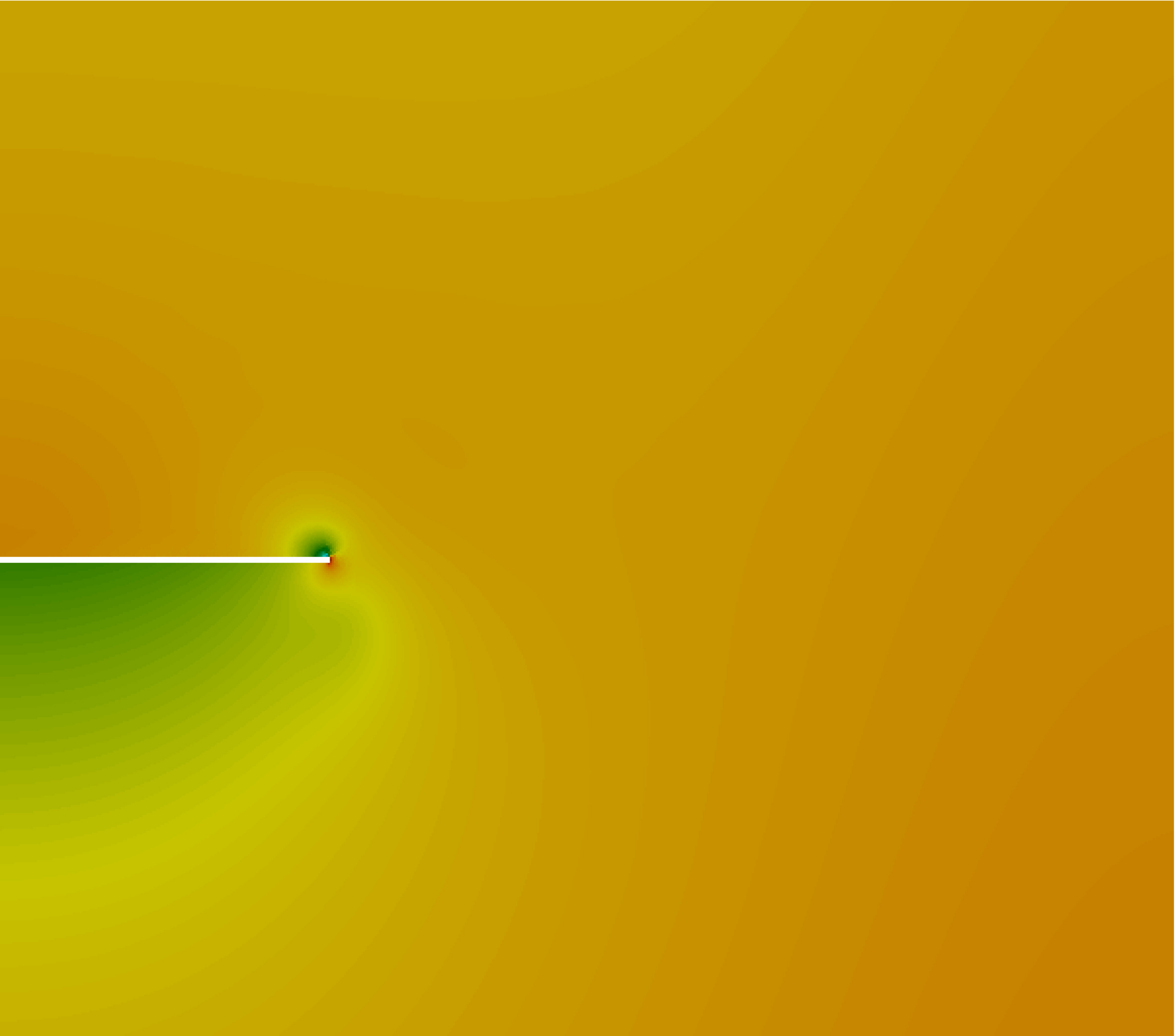}}
\subcaptionbox{}
{\includegraphics[width = 0.32\columnwidth]{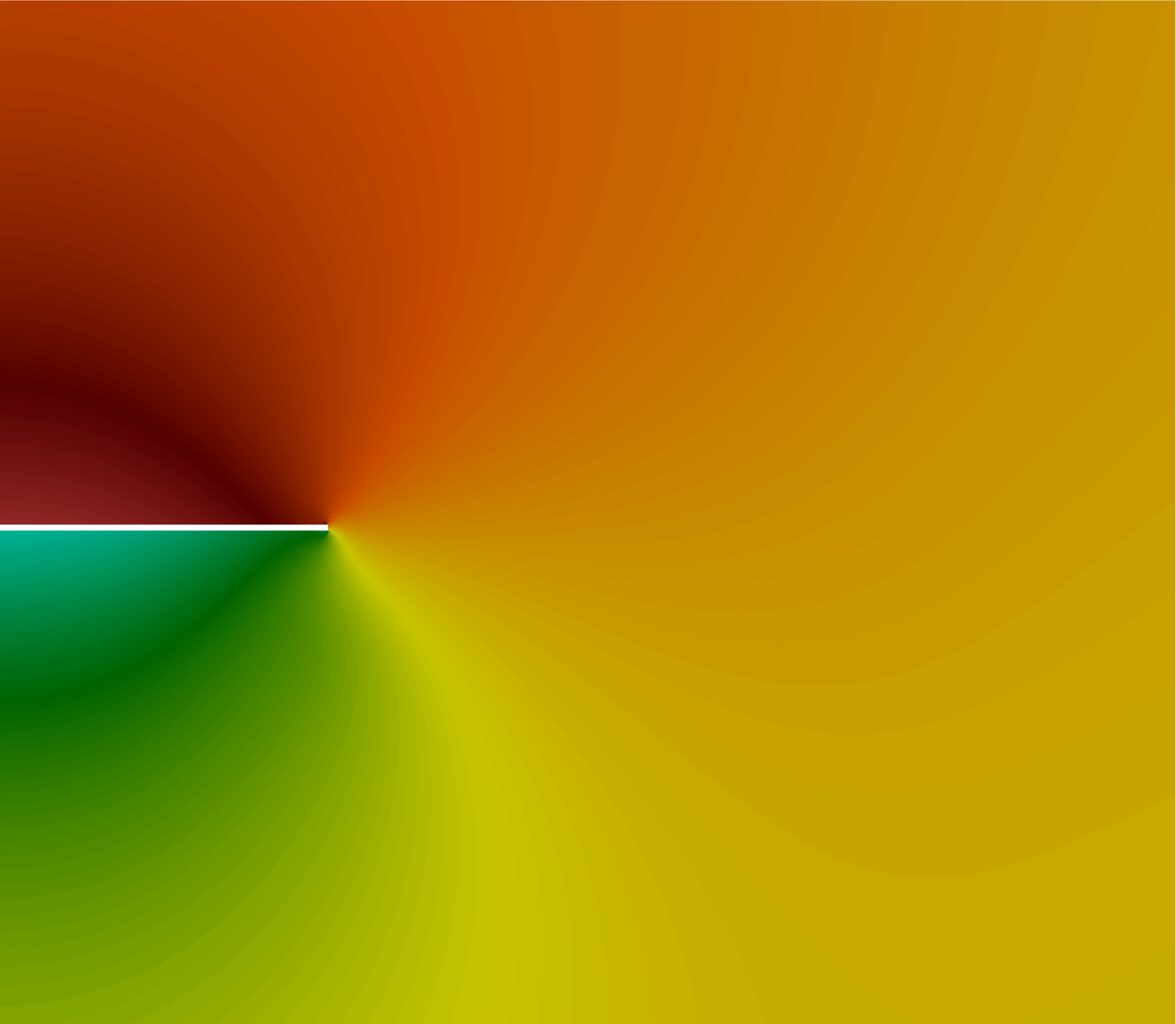}}
\caption{Detail of instantaneous velocity (a-c) and pressure (d-f) fields for $\beta=100$ and $\varepsilon=0.1$ at three different time instants. Only one symmetric half of the lamina is shown. Velocity and pressure are expressed in lattice units.
\label{fig:velpreFields}}
\end{figure}

We observe a good agreement with the considered literature, with the results obtained by the proposed method overlapping those available from previous works. In fact, the values for both components of the hydrodynamic function cast within similar ranges of those obtained by other numerical and experimental methods. In particular, our results suggest that variations of added mass effects are negligible for different values of the amplitude of oscillations. This is in contrast with findings obtained via standard LBM \cite{Falcucci}. Moreover, consistently with the results reported in the scientific literature, the dependence of the imaginary part on $\beta$ tends to reduce as $\varepsilon$ is increased.
\\
Finally, in Figure \ref{fig:velpreFields} we present the evolution of velocity and pressure fields, respectively, for a representative case of the fluid dynamic problem under analysis ($\beta = 100$, $\varepsilon = 0.1$). The fields are represented in lattice units.

The displayed frames in Figure \ref{fig:velpreFields} refer to three consecutive time instants, respectively. Overall, both velocity and pressure fields are qualitatively and quantitatively comparable with those presented in other flow visualization reported in literature. In particular, we note that the highest values of velocity are concentrated in proximity of the lamina tips with regions of null velocity occurring at the opposite side of the lamina (Figures \ref{fig:velpreFields}, frame (b)). Such structures indicate the formation of vortex shedding phenomena. This is consistent with what observed in \cite{AureliBasaran}. As far as the pressure field is concerned, we notice the presence of concentrated regions of relatively low pressure at lamina edges (Figures \ref{fig:velpreFields}, frame (d)). This is also noted in other works \cite{AureliBasaran,Falcucci,Abdelnour}.
Similar scenarios are observed for other sets of control parameters $\beta$ and $\varepsilon$ within the ranges analyzed.

\section{Numerical performance and implementation aspects}
\label{implementation}
In this section, we analyze the numerical performance of the moving grid approach for the hybrid lattice Boltzmann method and, finally, we discuss some implementation aspects.
\\
To asses the numerical efficiency of the MG-HLBM, we evaluate the computational cost required by the proposed method in comparison with the standard LBM. To this aim, we consider the simplified case of an hybrid mesh composed by the overlapping of a uniformly spaced structured region, which covers the entire computational domain, with the same unstructured mesh previously used in this study. In other words, we do not consider local refinements for the structured mesh.
In such a scenario and for the case of static unstructured mesh, we can apply the following equation, introduced in \cite{DiIlio}, to quantify the differences, in terms of computational cost, between hybrid and standard LB approaches:
\begin{equation}
G = \frac{T_\textbf{h}}{T_\textbf{eq}}=\frac{1+\chi s'n}{g}.
\label{eq:speedup}
\end{equation}
In equation \ref{eq:speedup}, $G$ is the speedup (gain) indicator, which is defined as the ratio of the wall-clock time per algorithm iteration associated to the hybrid method ($T_\textbf{h}$) over the wall-clock time per algorithm iteration with reference to an equivalent LBM implementation on uniform spaced mesh ($T_\textbf{eq}$), namely a standard LBM with similar resolution around the solid body.
A value of G less than 1 indicates that, by definition, the computational cost of the hybrid method is lower than that of standard LBM, when a similar resolution around the body is considered.
The speedup indicator $G$ depends on four parameters which are, respectively: the amount of unstructured nodes with respect to the number of structured ones ($\chi$), the ratio between the number of structured nodes which would be required by an equivalent LBM and the number of structured nodes actually used within the hybrid mesh ($g$), the ratio of the standard LBM processing speed to the specific processing speed associated to the unstructured method ($s'$), and the number of sub-iterations required by the unstructured method ($n$).
For the present case, the percentage number of unstructured nodes $\chi$ is equal to 0.244$\%$. The value of $\chi$ refers to the specific mesh under consideration. However, such a value is representative of a typical unstructured mesh required by the hybrid method and it is in line with those reported also in \cite{DiIlio,DiIlio2018}. In fact, we stress that the hybrid method requires only a small portion of the computational domain to be covered by the unstructured mesh, namely the region around the solid body. This represents one of the key elements of the proposed MG-HLBM.
Further, $n$ is set to 50. We emphasize here that the hybrid mesh adopted in this work has been designed on the basis of either a sensitivity analysis, as presented in Sec. \ref{Numerical setup}, and a stability analysis. As far as stability aspects are concerned, we recall that, due to the explicit time integration, both the standard LB and the unstructured method must satisfy the following condition: $\Delta t < 2 \tau$. While such a condition is always verified for the standard scheme, this is not necessarily the case for the unstructured one. This represents a clear limitation for the present hybrid method since it constrains, eventually, the value of $n$, namely the number of sub-iterations required by the unstructured method in order to synchronize the solution.
The parameter $s'$ was evaluated to be roughly equal to 4. Such an average value was obtained by actually measuring the code computational performance for several unstructured meshes. In addition, the number of nodes required by the equivalent standard LBM to have a similar resolution around the lamina is such that $g\simeq1.78$. With this set of parameters, we find  $G\simeq0.84$. This result indicates that, at each time-step, the hybrid method provides a gain in efficiency with respect to the standard LBM, when a similar resolution around the solid body is considered.
Notwithstanding a relatively low speedup, we emphasize that the code developed in this work still requires further optimization. Therefore, the value of G obtained for the specific case under analysis should be considered as an illustrative example.
\\
To complete the assessment of the MG-HLBM numerical performance we must include in the analysis the additional computational cost required by the mesh-motion and refill routine. In particular, such a routine is performed, at each time-step, by means of three sequential steps, that are: 1) motion of the unstructured mesh, 2) re-classification of interpolation nodes and donor elements, on structured and unstructured mesh, respectively, 3) refill procedure.
Despite this routine being laborious for the general case of arbitrary/induced motion of the unstructured mesh, its computational realization becomes particularly efficient when the motion of the unstructured mesh is imposed, such as in the present case, since it is known a-priori the location of the unstructured nodes at each time-step.
Another important aspect is related to the number of unstructured nodes. As observed also in \cite{DiIlio,DiIlio2018}, the number of unstructured nodes required by the hybrid method is significantly lower than the number of structured nodes used within the whole computational domain. For instance, in the present case, the number of unstructured nodes is less than 1$\%$ of the total number of structured nodes. This implies that, the interpolations needed to be performed within the refill procedure involve only a relatively low amount of computer operations.
In order to quantify the influence, in terms of computational performance, of the mesh-motion and refill procedure, we measured the computational time associated with such a routine. As a result, we found out that its processing speed is about two times higher than the processing speed associated to the unstructured method. Therefore, we can take into account of this additional computational cost by considering a modified number of unstructured sub-iterations $n'$ in equation \ref{eq:speedup}, such that: $n'$ = $n$ + 0.5. The overall computed value of the speedup indicator results then to be fairly the same of the one computed for the static case ($G\simeq0.84$). This is due to the fact that, the hybrid method, requires the unstructured routine to be sub-iterated several times within the same time-step. On the other hand, the moving grid routine is performed only one single time at the end of each time-step. For this reason, its computational cost results to be roughly two order of magnitude lower than the overall computational cost related to the unstructured routine. As a result, we can consider the computational time associated with the moving grid routine to be negligible with respect to the total.
In conclusion, in spite of a more complex algorithm, we observe that the numerical efficiency of the present MG-HLBM is, in general, comparable or even higher than the one associated to other approaches for moving objects based on standard LBM.

\section{Conclusion}
\label{Conclusion}
In this work, we extend the HLBM toward a moving grids approach and we explore its capability through the parametric study of a thin lamina undergoing transverse harmonic oscillations in a viscous quiescent fluid.
The numerical findings are in line with those presented in literature and obtained by different numerical strategies as well as experimental approaches. The performed analysis demonstrates that the proposed method is able to properly predict the main features of the fluid flow induced by the motion of the lamina. In particular, we observe a significant improvement, in terms of accuracy, with respect to previous works performed via standard lattice Boltzmann method on cartesian grid.
The proposed method represents a viable alternative also to other approaches such as the IB-LBM. In this regard, the MG-HLBM presents some key advantage. First of all, the hybrid strategy allows the possibility of applying an efficient local refinement. Second, the presence of an unstructured body-fitted mesh allows the direct evaluation of the forces at the real boundary of the body, thus leading to accurate results while preserving the computational efficiency of standard LBM.
These features make the MG-HLBM particularly appealing for multi-scale problems. On the contrary, the IB-LBM is based on a substantially different concept, therefore the choice between the use of this or the present method should be based, in general, on the peculiarity of the problem under analysis.
Also, the MG-HLBM and, in general, the hybrid method, can be regarded as a complex boundary condition for solid walls, which can be extended to deformable objects. This can be the subject of future works. These aspects are instrumental when dealing with fluid-structure interaction problems, involving complex geometries.

\section{Acknowledgements}
This work was supported by the Italian Ministry of Education, University and Research under PRIN grant No. 20154EHYW9 "Combined numerical and experimental methodology for fluid structure interaction in free surface flows under impulsive loading".
\\
The numerical simulations were performed on \textit{Zeus} HPC facility, at the University of Naples "Parthenope"; \textit{Zeus} HPC has been realized through the Italian Government Grant PAC01$\_$00119  \textit{MITO - Informazioni Multimediali per Oggetti Territoriali}, with Prof. Elio Jannelli as the Scientific Responsible.
\\
One of the authors (S. Succi) acknowledges funding from the European Research Council under the European Union’s Horizon 2020 Framework Programme (No. FP/2014-2020) ERC Grant Agreement No. 739964 (COPMAT).

\pagebreak


\end{document}